\documentclass[twocolumn,pre,showpacs]{revtex4}
\usepackage{epsfig}
\usepackage{amsfonts}
\usepackage{amsmath}
\usepackage{cancel}
\usepackage{graphicx}
\usepackage{mathcomp}
\usepackage{placeins}
\usepackage{floatflt}
\usepackage{float}
\usepackage{amssymb}
\usepackage{subfig}

\usepackage{times}

\newcommand{\IGN}[1]{}

\begin{document}
\title{Stabilization of biodiversity in the coevolutionary rock-paper-scissors
game on complex networks}
\author{Markus Sch\"utt and Jens Christian Claussen}
\affiliation{Institute for Neuro- and Bioinformatics, University of L\"ubeck, Ratzeburger Allee
160, D-23562 L\"ubeck, Germany}
%\affiliation{
%\thanks{Previous address: Institute for Theoretical Physics and Astrophysics, Christian-Albrecht-University Kiel, D-24098 Kiel, Germany}
\date{February 24, 2010}
%\date{\today}
\begin{abstract}
The dynamical mechanisms that can stabilize the coexistence of species
(or strategies) are of substantial interest for the
maintenance of biodiversity and in sociobehavioural dynamics.
We investigate the mean extinction time in the
coevolutionary dynamics of three cyclically
invading strategies for different evolutionary processes on various 
classes of complex networks, including random graphs,
scale-free and small world networks.
We find that scale-free and random graphs lead to a strong
stabilization of coexistence both for the Moran process and the
Local Update process. The stabilization is of an order of 
magnitude stronger compared to a lattice topology, and
is mainly caused by the degree heterogeneity of the graph.
However, evolutionary processes on graphs can be defined 
in many variants, and we show that in a process using effective payoffs
the effect of the network topology can be completely reversed.
Thus, stabilization of coexistence depends on both network geometry and
underlying evolutionary process.
\end{abstract}
\maketitle

\section{Introduction}
Dynamical mechanisms that stabilize cooperation have intrigued 
scientists from different fields for many decades,
and evolutionary game theory has developed from a 
coining metaphor to a well-established research area with
a wide range of applications from biology to behaviour
\cite{hofbauersigmund,gintis,nowakbook,szabofath,rocacuestasanchez}. 
An analogous question addresses the dynamical mechanisms
that can stabilize coexistence of (biological) species
or (behavioral) strategies \cite{loners}.
In a biological context:
 how is biodiversity sustained?
Here, cyclic coevolution has been suggested as to support 
the coexistence of strategies
\cite{sinervo,kerr,czaran}.
But it already has been reported that cyclic dominance alone is not enough to stabilize 
coexistence of strategies \cite{kerr,E. coli Raum 3}.
While space (or, in ecological setting, niches) 
provides a means of stabilizing biodiversity, it is not the only 
mechanism to be taken into account.
One additional way of stabilization is the introduction of a non-zero 
sum game, which also can stabilize finite but sufficiently large
populations \cite{CT08} and 
results in an 
exponential scaling of the mean first extinction time \cite{unstrukturiert}.

In general, the stability of the coexistence fixed point
depends on several instances:
firstly, on the payoff matrix \cite{czaran,CT08},
secondly, on the population size and the underlying
evolutionary process \cite{TCH05,TCH06}, and thirdly, 
on the spatial structure of the population which
can assume interaction topologies with the structure
of a lattice or more general graphs.
A spatial discretization of
 coevolutionary dynamics
can lead to a stabilization of the coexistence of game theoretical strategies, 
compared to well mixed populations
\cite{kerr,czaran}. 
This corresponds to the 
effect of spatial discretization in 2-strategy games,
where 
-- despite exceptions
\cite{hauertdoebeli} --
cooperative strategies can be stabilized
\cite{lindgren94,herz94}.
In this direction, many works deal with structured 
versions of the prisoners dilemma \cite{prisoners}, 
where it has been shown that 
cooperators can coexist with defectors even in parameter regions where cooperation will never be observed in mixed populations 
 \cite{pioneering work,similarity,pioneering work 2,szolnoki,myopic,szabotoke98,szabohauert02,butlergoldenfeld09}. 
In contrast to these two-strategy evolutionary games, 
here we analyze the widely known rock-paper-scissors game (RPS) 
which includes three strategies with cyclic dominance. 
Such three-strategy models including cyclic dominance are well-known in game theory and related fields 
\cite{marine cyclic evolution,vertebraten,sinervo,eidechsen2,kerr}.
In a spatially extended system, 
the coexistence of the three cyclically invading strategies
 is stabilized,
even in cases where it is not stabilized in the well-mixed system 
\cite{czaran}.
--
But how strong is this stabilization? Are there differences between the stabilization effect on different graphs and for different processes? How does the population size influence the stabilization? 
To answer these questions, 
in this paper 
we examine 
three different update mechanisms 
-- the frequency-dependent Moran process 
\cite{moran,original_Moran}, the local update process \cite{TCH05}, 
and a process adapted from 
Szolnoki, Perc, and Danku \cite{szolnoki} -- 
and a broad range of complex network types as underlying structures, 
and,
for each network type, analyze the
mean extinction time (\textit{MET}),
i.e.\ the 
average time until one of the three strategies has gone extinct.

\section{The rock-paper-scissors game}
We consider a population of $N$ individuals 
or agents. $N$ is assumed to be constant all the time
(although this common assumption is an approximation,
and additional effects emerge, e.g.,
in growing populations
\cite{poncela09}).
Each individual is placed on one vertex of a graph containing $N$ vertices 
and can take on one of the three strategies 'rock', 'paper', or 'scissors'. 
As played on schoolyards, 'rock' looses against 'paper' but wins against 'scissors', 
and cyclically permutated. 
In game theory, the `interaction kernel' quantifying the outcome of 
agent collisions 
is cast into a payoff matrix,
which reads for RPS
\begin{equation}
P
=\left(\begin{array}{ccc}
0&-1&1\\1&0&-1\\-1&1&0
\end{array}\right),
\end{equation}
and we assume that 
 we have a zero-sum RPS game. We now can find the total payoff for a certain individual 
$i$ playing strategy $S(i)$ by adding all payoffs that $i$ gets from neighbored nodes. 
With 
$L$ 
as the adjacency matrix (whose elements $L_{i,j}$ are $1$ if two vertices $i$ and $j$ are connected by a link, and $0$
otherwise),
we can compute the 
total payoff $\pi_i$ for an individual $i$ by
\begin{equation}
\pi_i=\sum_{
j=1,j\neq i
}^{\scriptsize N}L_{i,j}P_{S(i),S(j)}.
\end{equation}
Assume the strategy $S(i)$ to be $\alpha$. 
Let us use the notation $\beta$ for the strategy 
dominated by $\alpha$, and $\gamma$ for the strategy that dominates $\alpha$, with $\{0,1,2\}$ as numerical values that represent these strategies. 
Then 
using the special structure of our payoff matrix, 
this term can be simplified to
\begin{equation}
\pi_i=\sum_{\scriptsize j=1}^{\scriptsize
N}L_{i,j}\left(\delta_{S(j),\beta}-\delta_{S(j),\gamma}\right).
\end{equation}
This total payoff can be modified for normalization issues, as we will describe in the next section, because the way of normalization 
varies in the different update mechanisms,
defining rules after which strategy changes occur, 
depending on $\pi_i$. Note that this total payoff can vary relevantly from the case of unstructured populations. For example, consider the case of a node $i$ playing strategy $\alpha$. 
Let $j$ be the only vertex $i$ is connected to, and let $j$ be the only vertex that plays strategy $\gamma$ which may dominate $\alpha$ so that $i$ gets a total payoff $\pi_i=-1$ 
although in the whole population there is only one agent that dominates strategy $\alpha$. 
This 
single agent playing strategy $\gamma$
would only have negligible influence on the payoff of $i$ in a well mixed population, namely, reduce it by $1/(N-1)$.

\section{Update mechanisms}
\subsection{The frequency-dependent Moran process on networks}
\vspace*{-2ex}
For the first update rule we have adapted the well known frequency-dependent \cite{moran} Moran process \cite{original_Moran}. 
For normalization, we divide the total payoff $\pi_i$ 
by $k(i)$ (being the vertex degree, number of neighbors of $i$). 
So the payoff reads
\begin{equation}
\pi_i=\frac{1}{h(i)}\sum_{\scriptsize j=1}^{\scriptsize
N}L_{i,j}\left(\delta_{S(j),\beta}-\delta_{S(j),\gamma}\right).
\end{equation}
Now we choose a vertex $i$ and another vertex $j$, for which $L_{i,j}=1$ holds, 
at random. With probability
\begin{equation}
\phi_M^{S(j)\to S(i)}=
\frac{1}{2}
\frac{1-\omega+\omega \pi_i}{1-\omega+\omega\langle\pi\rangle}
\end{equation}
the individual on the vertex $i$ reproduces, and its strategy replaces the strategy of the player on the vertex $j$. Here $\omega$ is the strength of selection and
\begin{equation}
\langle \pi\rangle=\sum_{j=1}^N L_{i,j}\pi_j/k(i)
\end{equation}
is the mean fitness of the relevant subpopulation, 
in our case,
all neighbors of $i$. Note that in contrast to the Moran process in well mixed populations, $\langle\pi\rangle$ needs not to be $0$ even for the case of zero-sum RPS we have analyzed here (although this still holds for the 
expectation value $\langle\pi\rangle$). 
Note also that the original definition of the Moran process is 
slightly different because one should choose an agent $i$ for reproduction at random proportional to $\phi_M$. That means we are ensured that there is a real update event in every update step, 
which is not the case in our definition.
Our definition is numerical cheaper because it is not necessary to compute the payoffs for all of the vertices in every time step, and it influences the time scale only by a constant factor.

\subsection{Local update process on networks}
\vspace*{-2ex}
As the availability of the global information $\langle \pi \rangle$ to all
individuals 
may often be unrealistic, one should also consider processes where only
locally available information is of relevance to the dynamics.
In the extreme case, only 
randomly selected pairs of nodes compete,
as in the class of local comparison processes
\cite{helbing92,helbing93,blume93,szabotoke98,hauertszabo05,C07}.
From this class, 
we have investigated the local update process \cite{TCH05} as well. 
Here we, again, normalize the payoff of a vertex $i$ by dividing it by the number of its neighbors. Then we again choose two linked vertices $i$ and $j$. With probability
\begin{equation}
\phi_{LU}^{S(j)\to S(i)}=\frac{1}{2}+\frac{\omega (\pi_{a}-\pi_{b})}{2\Delta\pi_{max}}
\end{equation}
$j$ changes its strategy to the strategy of $i$. Here, $\Delta\pi_{max}$ is the greatest possible payoff difference, that is the difference between the largest and the smallest entry in the payoff matrix.

\subsection{SPD-process}
\vspace*{-2ex}
As a third update rule we have modified a process introduced by 
Szolnoki, Perc, and Danku \cite{szolnoki}  
which one could think of as a simplified version 
of the local update process at the first view. 
We choose a vertex $i$ per random and another vertex $j$ which is connected to $i$ by a link also per random. In contrast to the Moran and the local update process, both payoffs are modified according to the rule
\begin{equation}
\overline{\pi_i}=a\pi_i+(1-a)\frac{\pi_i}{k(i)}.
\end{equation}
(In the case $a=0$ we again have the payoffs like in the other two processes.) Now, only if $\overline{\pi_j}>\overline{\pi_i}$, $j$ reproduces with probability
\begin{equation}
\phi_{SPD}^{S(i)\rightarrow S(j)}=(\overline{\pi_j}-\overline{\pi_i})/\max(\overline{\pi_j}-\overline{\pi_i})
\end{equation}
proportional to the difference of the payoffs, and the strategy of vertex $j$ replaces the strategy of vertex $i$. The maximum in this term has to be understood as the the maximum of the 
possible payoff difference that two vertices with degrees $k(i)$ and $k(j)$ can have. This ensures the probability to always be in the range of $0$ and $1$. One can easily compute this maximum, so we can achieve the following:
\begin{equation}
\phi_S^{S(i)\rightarrow
S(j)}=\frac{\overline{\pi_j}-\overline{\pi_i}}{2(1+a(k_{max}-1))}.
\end{equation}
Here $k_{max}$ is the maximum of $k(i)$ and $k(j)$.
\vspace*{-2ex}
%\clearpage

%\begin{widetext}
\begin{figure*}[htbp]
\subfloat[]{\epsfig{file=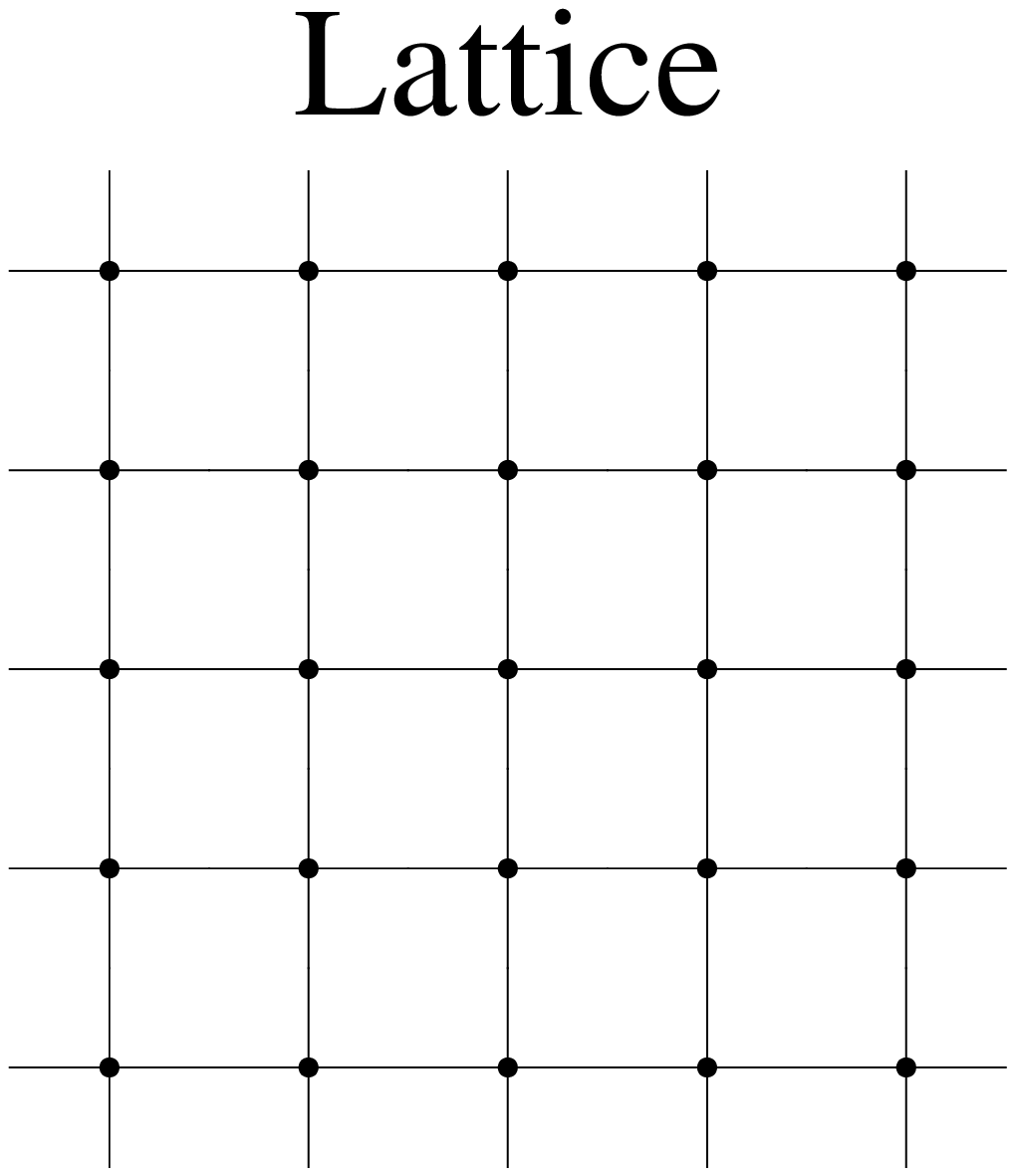,height=32mm}}
~~~~~~~~
\subfloat[]{\epsfig{file=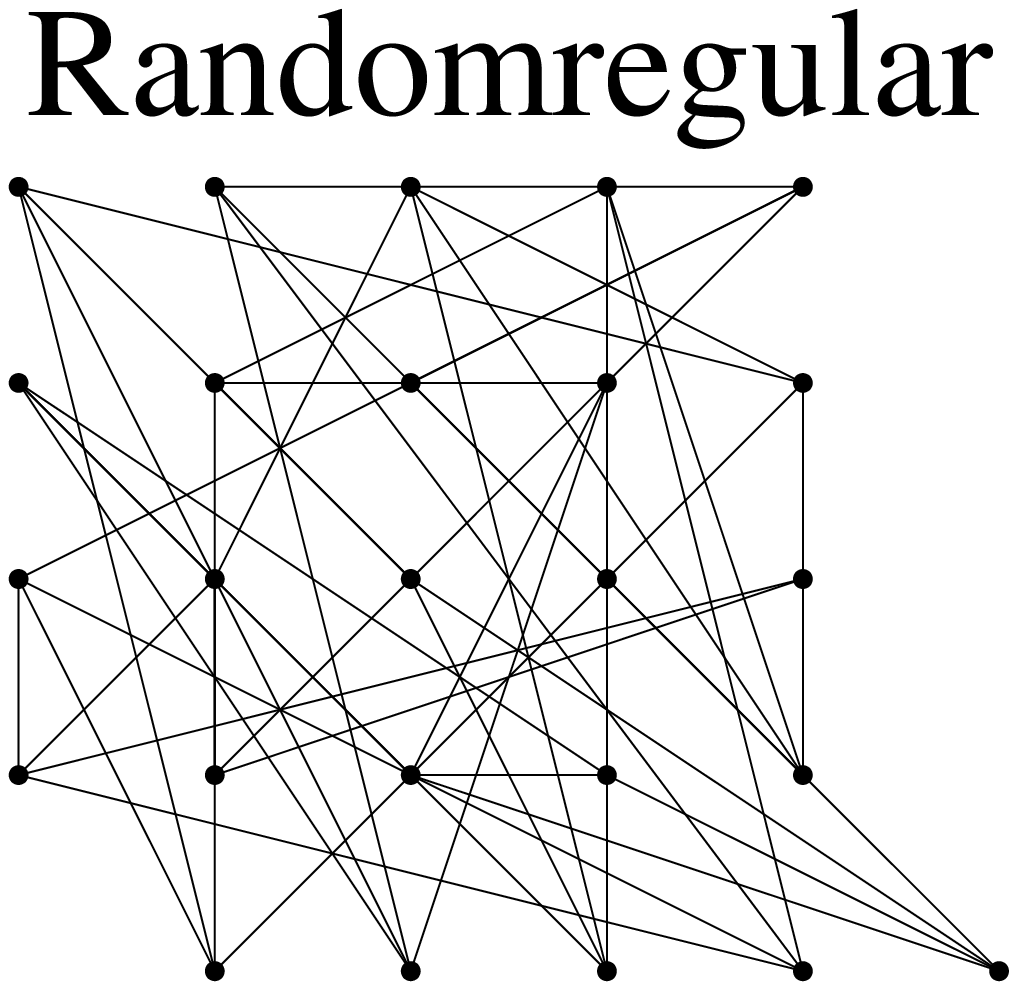,height=32mm}}
~~~~~~~~
\subfloat[]{\epsfig{file=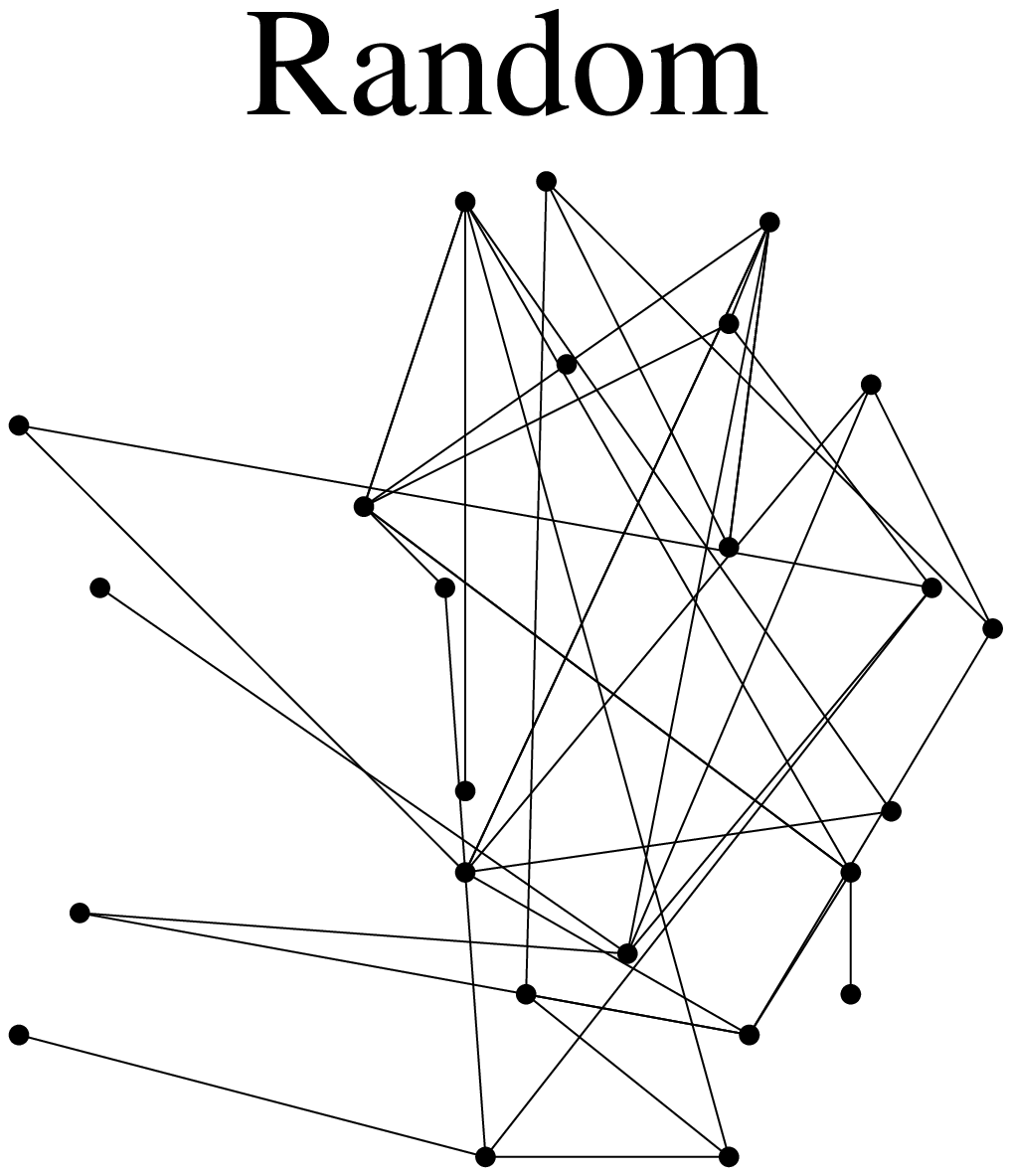,height=32mm}}
~~~~~~~~
\subfloat[]{\epsfig{file=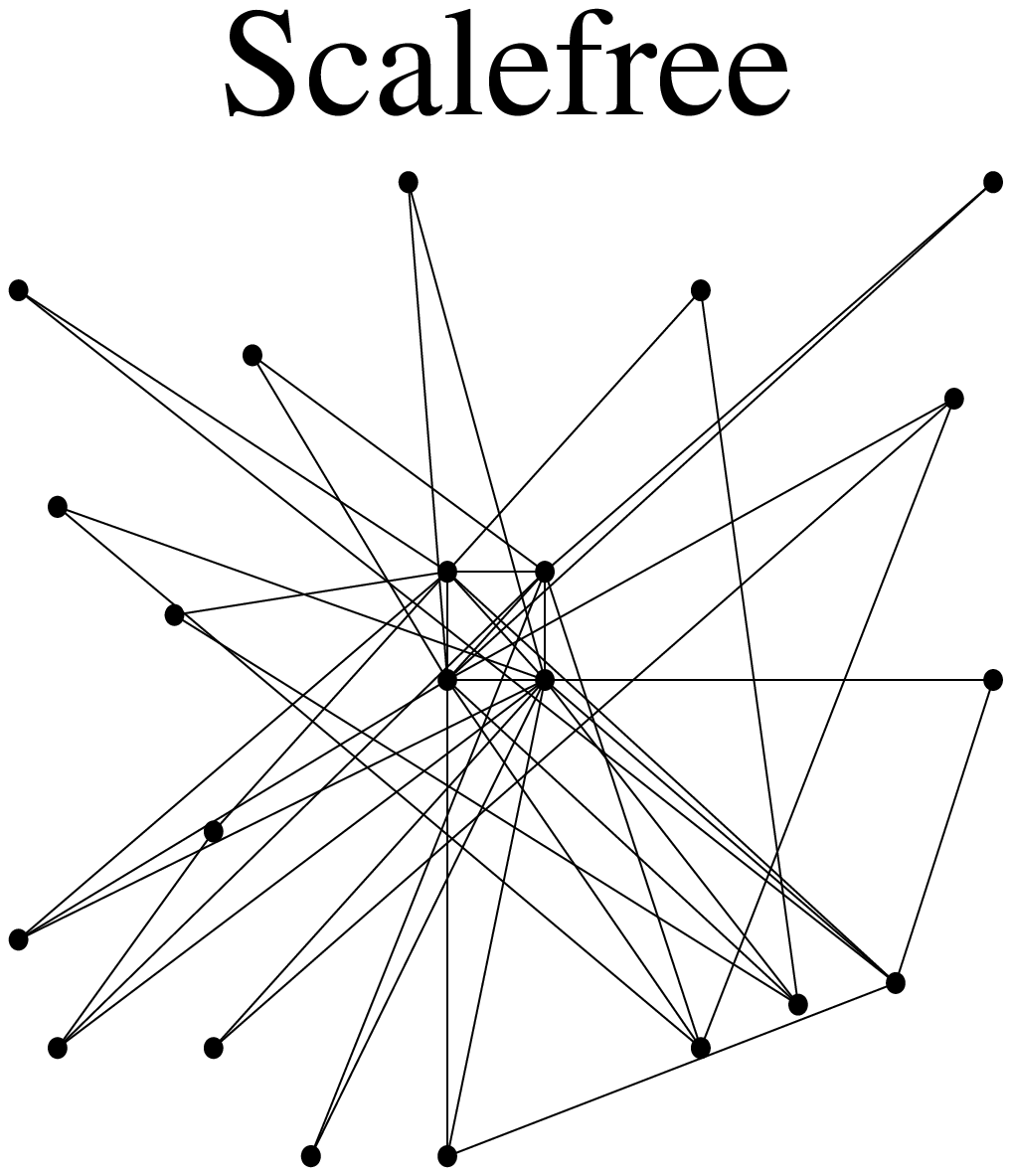,height=32mm}}
~~~~~~~~
\subfloat[]{\epsfig{file=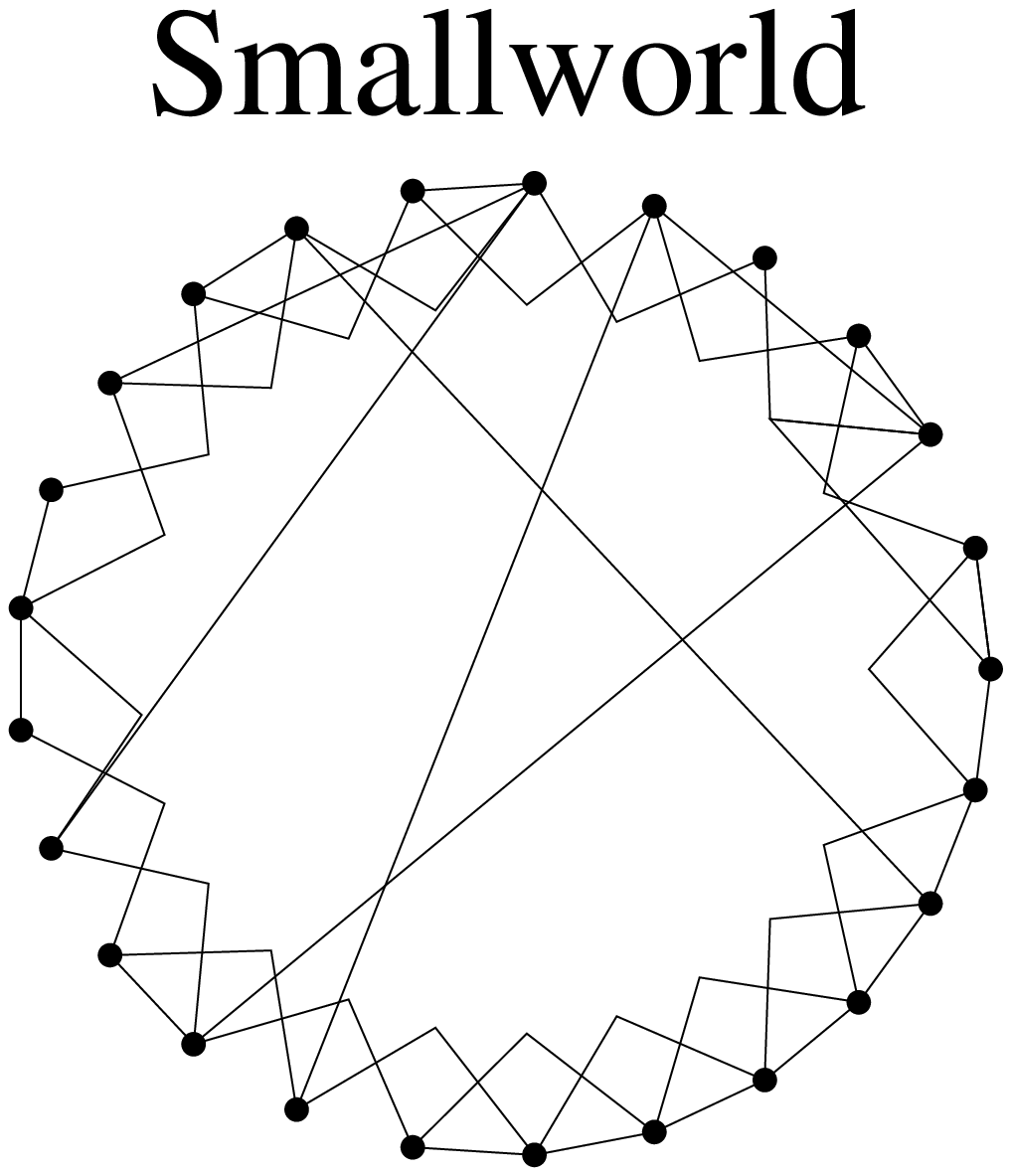,height=32mm}}
\\
\noindent
\begin{minipage}{\textwidth}
\noindent
\caption{Realizations of five different types of networks with an average vertex degree of 
$\overline{k}=4$ and a total number of vertices of $N=25$. (a) Square lattice with periodic constraints. The half-links are clear illustrations of the periodic constraints. (b) Random regular graph. Each vertex has the same degree 
$\overline{k}$. (c) A connected ER-random graph. (d) Scalefree network after 
Barab\'asi and Albert. The nodes placed on the square in the middle are the 
$k_0$ starting nodes. (e) Smallworld network after Watts and Strogatz with $p=0.25$.
}
\label{Graphendarstellung}
\end{minipage}
\end{figure*}
%\end{widetext}

\section{Investigated networks}
All networks used as an underlying topology for such a game should fulfill some 
common properties.
They should be connected, so that we do not have some parts of the population not 
being linked to the rest of the population. 
Here, we always consider undirected and unweighted (binary) graphs.
To avoid self-interactions of the agents, 
self-loops should be excluded. 
A schematic illustration of the network types 
used in this paper 
is shown in
Fig.~\ref{Graphendarstellung}.

\subsection{Regular lattice}
The first step of complexity
of placing vertices
spatially 
-- beyond the linear chain --
is a regular square (or rectangular) lattice.
Typically, each node is connected with $4$ or $8$ nearest neighbors. To eliminate boundary effects, we use periodic boundary conditions. This is a very simple, but in many cases not realistic model, so we have analyzed some other specific types of models that will be presented in the following subsections. For a better comparability with the regular square lattice with natural connectivity of $4$ or $8$, we have restricted ourselves to graphs with an average vertex 
degree of $4$ and $8$.

\subsection{Erd\H{o}s-Renyi random graph}

The most frequently used model of real, not lattice-like graphs in the past decades was the 
Erd\H{o}s-Renyi (ER) random graph  \cite{Erdos}. Here one starts with a graph with $N$ nodes and connects each pair of nodes with 
probability $p$. In this model the probability of finding a node with degree
$k$ follows a Poisson distribution
\begin{equation}
P(k)=\frac{e^{-\lambda}\lambda^k}{k!},
\end{equation}
where
\begin{equation}
\lambda=N\left(\begin{array}{c}N-1\\k\end{array}\right)p^k(1-p)^{N-1-k}.
\end{equation}
The graphs we have investigated are constrained in two ways. On the one hand, they must be connected, what means that there is at least one possibly path from each node to any other so that we don't have two or more groups of not interrelated agents. On the other hand, we restrict ourselves to networks with a mean vertex 
degree of $4$ or $8$ for better comparability with the regular lattice. Because we will need only connected graphs with a certain mean 
degree $\overline{k}$, 
and because we would find these networks only with some 
probability even by adjusting $p$ with respect to the number of nodes, we have used the following
algorithm:
\\
\begin{enumerate}
\item Start with a graph \textbf{G} with $N$ vertices and no edges.
\item Link randomly chosen vertex pairs if they are not already linked until the total number of edges equals 
$N\overline{k}$. Now the mean vertex degree will equal $\overline{k}$.
\item If the graph is connected, accept it, otherwise return to step (1).
\end{enumerate}

\subsection{Random regular graph}

A compromise between the random graph and the regular lattice is the random regular graph with the same 
degree $\overline{k}$ for all vertices but a much shorter average shortest path length than in a lattice. For generating the random regular graphs we have used an algorithm proposed by Steger and Wormald \cite{randomregular}. This algorithm ensures to give connected random graphs with equal 
degree $\overline{k}$ for all vertices:
\\
\begin{enumerate}
\item Start with a graph \textbf{G} with $N$ nodes $\left (1,2,...,n\right)$ and no edges.
\item Repeat the following until the set \textbf{S} is empty: Let \textbf{S} be the set of vertex pairs $\{u,v\}$ of \textbf{G} that are not connected by an edge yet with both having at most 
degree $\overline{k}-1$. Then choose a specific pair $\{u,v\}$ out of \textbf{S} with probability proportional to
$\left(\overline{k}-k(u)\right)\left(\overline{k}-k(v)\right)$, where $k(u)$ and
$k(v)$ are the degrees of $u$ and $v$ in the graph generated up to now.
Add the edge to \textbf{G}. Delete the pair $\{u,v\}$ from \textbf{S}.
\item If \textbf{G} is $\overline{k}$-regular (that means connected and of the same 
degree $\overline{k}$ for all vertices), accept it, otherwise return to step (1).
\end{enumerate}

The last step with the return statement is necessary because unfortunately 
the algorithm does not exclude the possibility of receiving disaggregated 
graphs 
(which occurs rarely) 
or that the last connection possibilities are only self-loops.

\subsection{Barab\'asi-Albert scalefree network}

Many real networks, primarily huge ones, are described by the ER-model in an insufficient way. In many of these networks, 
for instance the
internet \cite{faloutsos3},
the WWW \cite{www1,www2}, the network of power nodes in the west of the United States \cite{small world}, the network of scientific collaborations 
\cite{paper}, or the metabolic network of yeast \cite{Hefe} one observes a 
degree distribution that (approximately) follows a power law,
\begin{equation}
P(k)\propto k^{-\gamma}.
\end{equation}
Barab\'asi and Albert \cite{scalefree} have proposed the following algorithm of modelling complex real networks:

\begin{enumerate}
 \item Start with a small number $k_0$ vertices that are all connected to each other.
\item Add another node and link it to an already present vertex with probability proportional to the degree of this vertex. Add further links in the same fashion until the new vertex has a degree 
$k<k_0$.
\item Repeat step (2) (until you have reached some value of $N$ you want).
\end{enumerate}

In contrast to the ER-model for random graphs or the likewise frequently used Watts-Strogatz-Modell (\cite{small world}, 
see next subsection) this model incorporates two main aspects of real networks, 
growth 
and preferential attachment. Namely, most real networks do not have a 
fixed size but grow like the WWW \cite{www1,www2}. 
The WWW is also an example for the preferential attachment: When creating a new website, one will typically try to link it to well known and popular sites to gain more attention for ones own site. Because of that, the 
Barab\'asi-Albert network has a high density of so-called hubs. These are vertices that are connected with many other nodes with relatively small vertex degree.

\subsection{Small-World-Network after Watts and Strogatz}

Another frequently used network model is the Small-World network. 
The existence of long range connections with simultaneous appearance of a high local cluster coefficient is a phenomenon often obtained in real networks \cite{small world 2}.). The notion integrates some slightly different types of networks. 
Here we have used the original model of Watts and Strogatz \cite{small world}. 
In their model one starts with a circle of nodes. All of them are connected with 
$k$ nearest neighbors ($k/2$ in the one direction and $k/2$ in the other). 
Afterwards, one starts with rewiring the links, beginning with an arbitrary vertex $i$ on the circle. 
With a certain 
probability $p$, $i$ is disconnected from its nearest neighbor in clockwise 
direction and
is connected to a randomly chosen vertex $j$ that is not already linked to $i$. In this way one goes on in clockwise direction until all links to nearest neighbors have been checked for rewiring. Then one goes on with the second nearest neighbor, and so on, until all links have been checked for rewiring.

For $p\to 0$ one achieves obviously some kind of regular lattice (that is structured in a different way that the one we have used). In contrast to that, for $p\to 1$ one achieves a graph equivalent to the ER-random graph. 
But for intermediate values of $p$ one gets networks with high local cluster coefficients like a regular graph (or lattice), but also a small average shortest path length like a 
random graph \cite{albertbarabasi}. 
For the sake of clarity, 
we have used $p=0.25$ in all cases.

\subsection{Network with Gaussian degree distribution \label{stubs}}

As a sixth type of networks, we have used a graph with a Gaussian degree distribution. This type is a compromise between the ER-random graph and the random regular graph. To generate these networks we have modified the algorithm used for the random regular graph by making the 
designated vertex degree $k$ dependent of $i$, $i={1,2,...,n}$. 
For these $k(i)$ we have used a Gaussian distribution with a 
standard deviation of 
${\overline{k}}/{3}$,
where $\overline{k}$ is the designated average vertex degree. By this choice of the standard deviation it is ensured that almost all designated degrees 
$k(i)$ are positive. We have ignored the few occurring non-positive (negative or $0$) 
degrees by choosing a new designated degree in the case of such a choice.

\subsection{Uncorrelated scalefree graph}

To analyze possible effects of the vertex degree distribution in scalefree
graphs after 
Barab\'asi and Albert on the dynamics of our games, we have investigated the dynamics on uncorrelated scalefree graphs as well. For the construction of such networks we have used the algorithm
described in the preceding subsection (\ref{stubs})
 with a power law degree distribution identical with the distribution of vertex degrees of the 
Barab\'asi-Albert scalefree graph instead of a Gaussian distribution. 
To avoid confusion, we will refer to this graph short as 
\textit{uncorrelated}. 
If we refer to scalefree graphs 
(without mentioning anything about the correlation) 
 we always consider the 
Barab\'asi-Albert network.

\section{Mean extinction times}

We have carried out extensive simulations to obtain the mean time until one of the three strategies has gone extinct when starting with each of the strategies having the same frequency $N/3$ (this requires that $N$ is an integer multiple of three). 
After the first strategy has vanished, 
the game lacks its cyclic dominance, and we return to the two strategy case 
which is well known and 
will 
end up
in a much shorter time
in a single-strategy steady-state because one of the two remaining strategies definitely dominates the second. Note that we measure time in units of single update steps instead of 
Monte Carlo steps, where on average each agent has been updated in one time step, so there is a factor of $N$ between these two time scales.

\subsection{Mean extinction time in well mixed populations}

First, let us consider the case of well mixed populations for 
comparison, which is identical to the case of fully connected graphs 
(that means each vertex is directly linked to any of the $N-1$ others). 
We have to distinguish between two cases: 
the Moran and the local update process on the one hand, and the SPD-process on the other hand.

\subsubsection{Moran and local update process in mixed populations}
 In well-mixed populations, the 
Moran and local update process
have been shown in 
\cite{TCH05}
 to have the adjusted replicator equation \cite{adjusted replicator equation} and the standard replicator equation \cite{hofbauersigmund} as limits as $N$ goes to infinity. They both have a neutrally stable (for the zero-sum RPS) fixed point in the state space of strategy densities at $\left(\frac{N}{3},\frac{N}{3},\frac{N}{3}\right)$. For any other point inside of the simplex, both replicator equations predict neutrally stable oscillations around this fixed point, but in the case of finite populations the population can run out of the fixed point and the orbits around it because of stochastical fluctuations, and end up on the edge of the simplex $S_3$ which is the boundary of the state space. A point on the edge of this simplex corresponds to the extinction of at least one of the three strategies (there are three points on the boundary corresponding to the survival of only one strategy). So we expect the MET of this system to tend to infinity as $N\to\infty$. But at this point there are still two open questions: 
What is the exact scaling dependence of the MET on $N$?
And how does the selection strength $\omega$ influence this dependence?

Our simulations give clear and short answers for both questions: In the zero-sum RPS game, the MET is 
proportional to $N^2$ 
(in single birth-death update steps, 
corresponding to $N$ Monte Carlo steps),
and independent of $\omega$ (both is different in the non zero-sum RPS case). For both processes, Moran and local update process as well, 
we have found 
the mean extinction time $t_{ext}$ to scale as
\cite{unstrukturiert}
\begin{equation}
\langle t_{ext}\rangle=0.54(\pm0.02)N^2.
\end{equation}

\subsubsection{SPD-process in mixed populations}
As this process has never been defined for unstructured populations let us use the complete graph as a model for the mixed population for simplicity, so that we 
do not have the risk of making any alterations to this update rule if re-defining it. Now we find that, when starting with $\left(\frac{N}{3},\frac{N}{3},\frac{N}{3}\right)$, each node has as many neighbors of the dominating strategy as of the dominated one, namely $N/3$. 
Because of the structure of the payoff matrix the fitness is identical to zero for 
every strategy and node. So the condition that the second chosen vertex has a greater fitness than the first is never fulfilled, and one will never observe a change in the strategy densities. For this reason the MET becomes infinite even in finite populations (at least when starting in the point $\left(\frac{N}{3},\frac{N}{3},\frac{N}{3}\right)$; even when starting in some neighboured point in the state space we would only achieve some conditional mean extinction time).

\subsection{Moran process on networks}

\begin{figure*}[pthb]
\epsfig{file=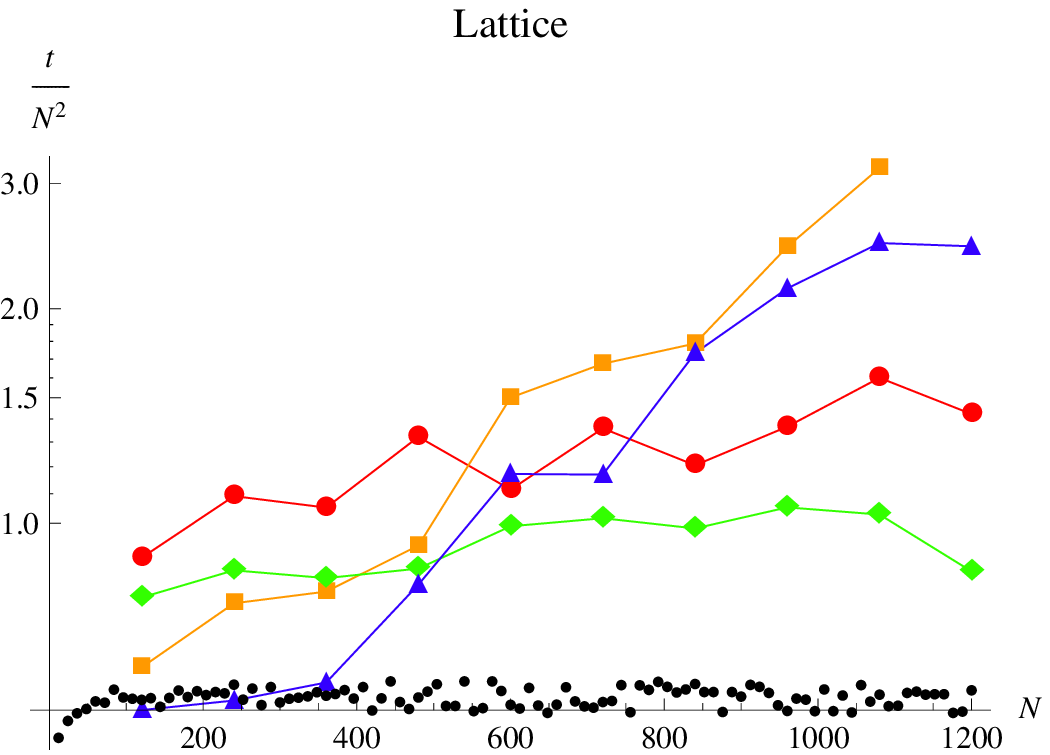,height=50mm}
\epsfig{file=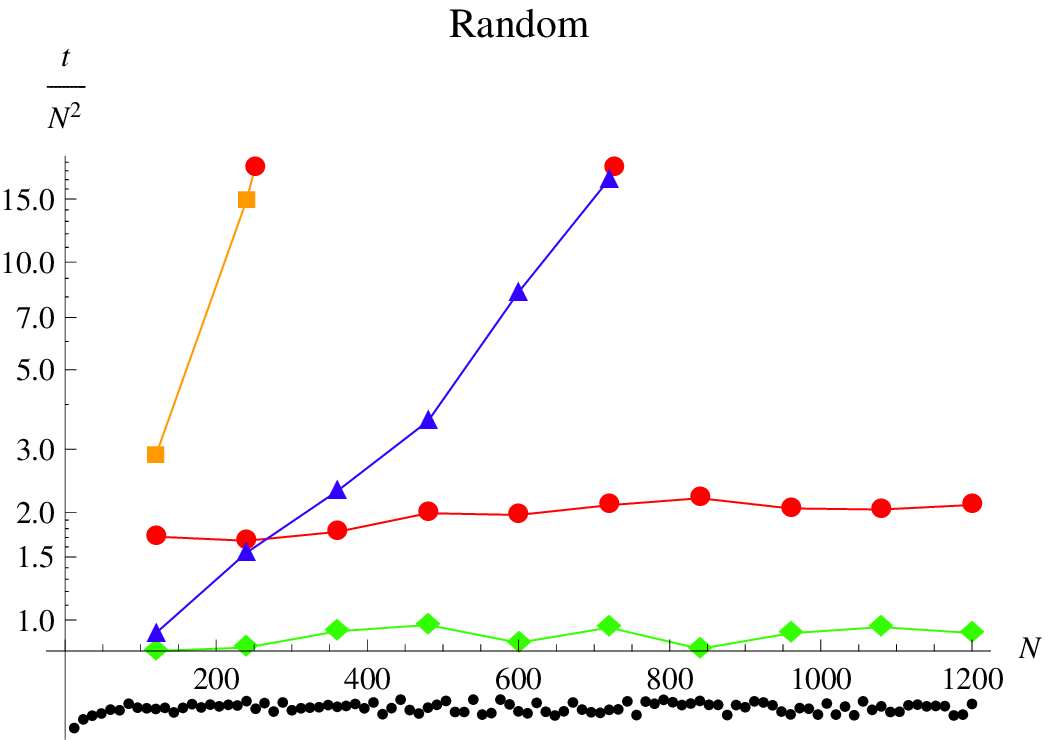,height=50mm}\\
\epsfig{file=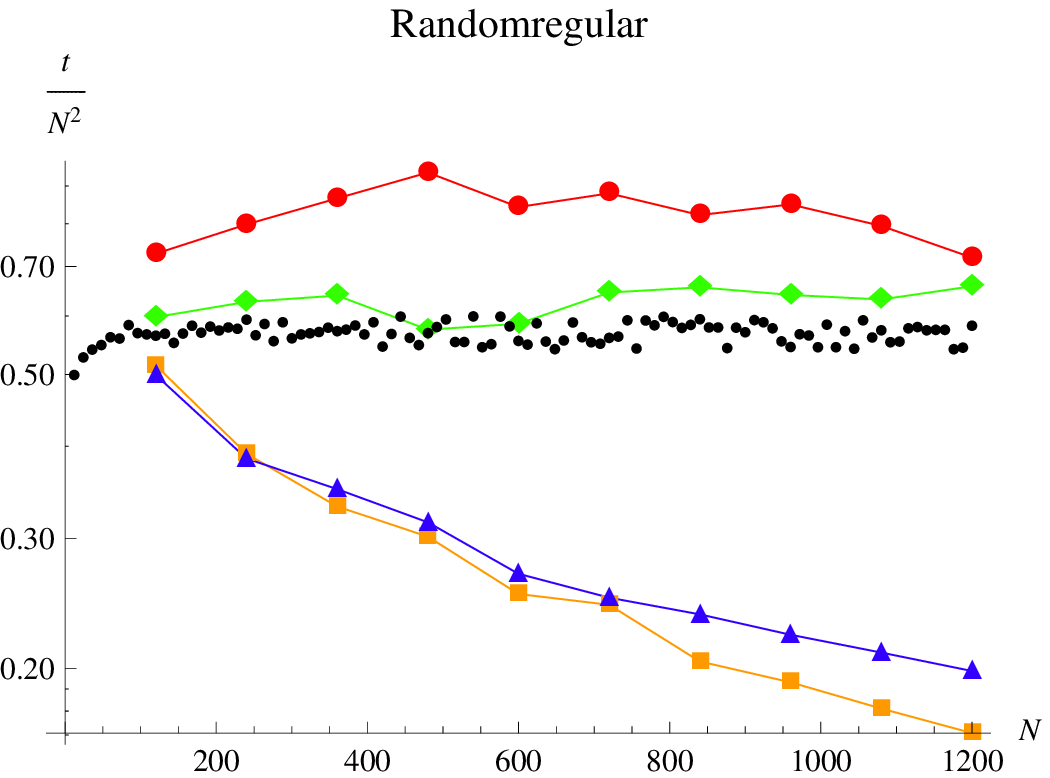,height=50mm}
\epsfig{file=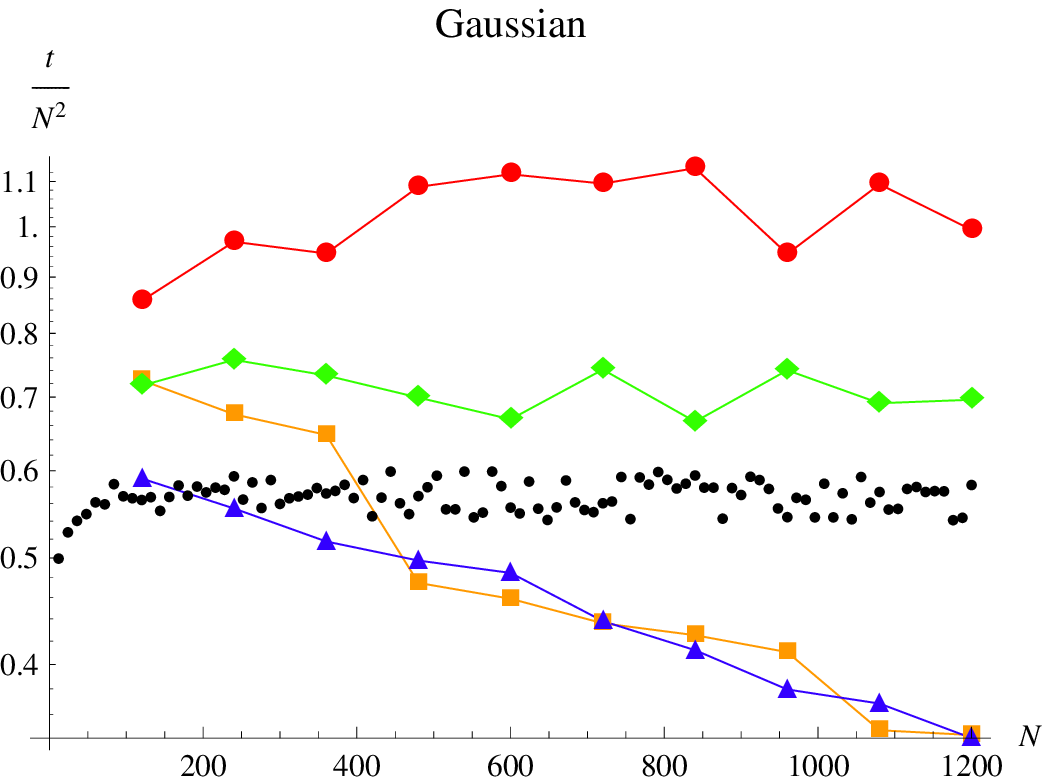,height=50mm}\\
\epsfig{file=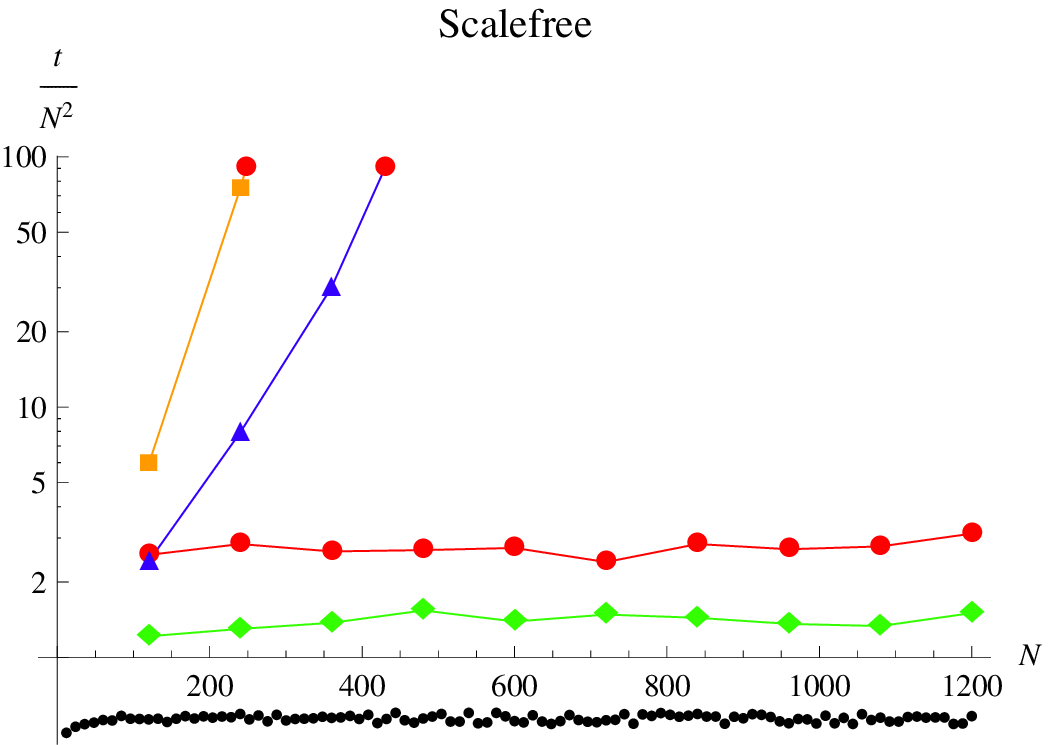,height=50mm}
\epsfig{file=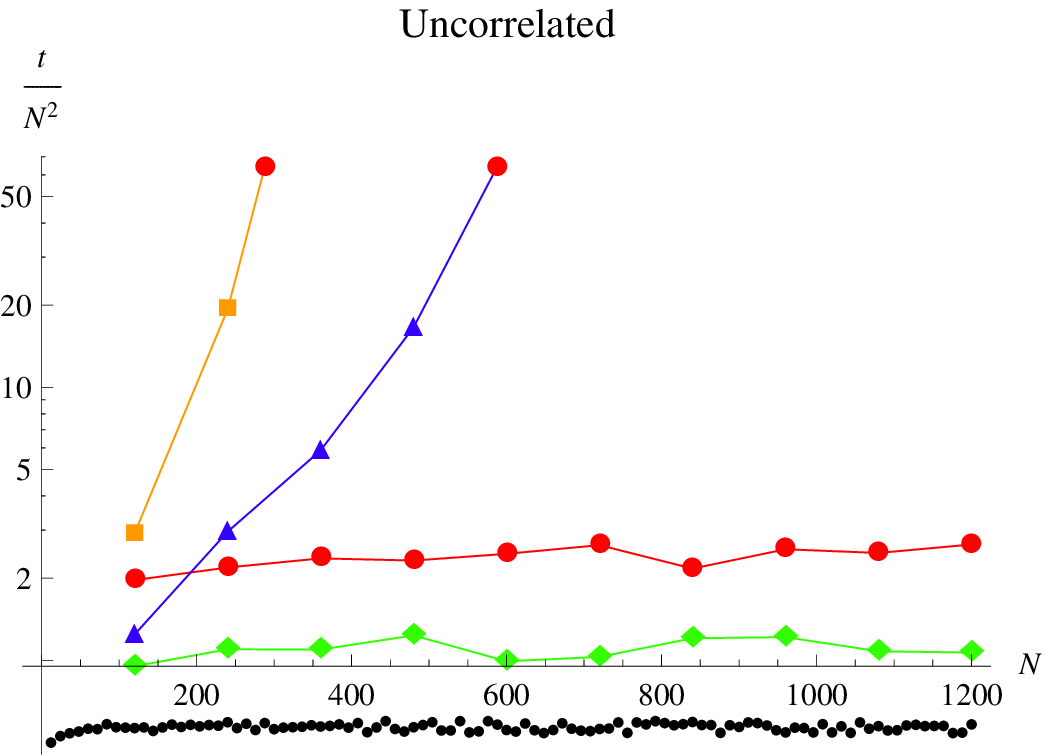,height=50mm}\\
\epsfig{file=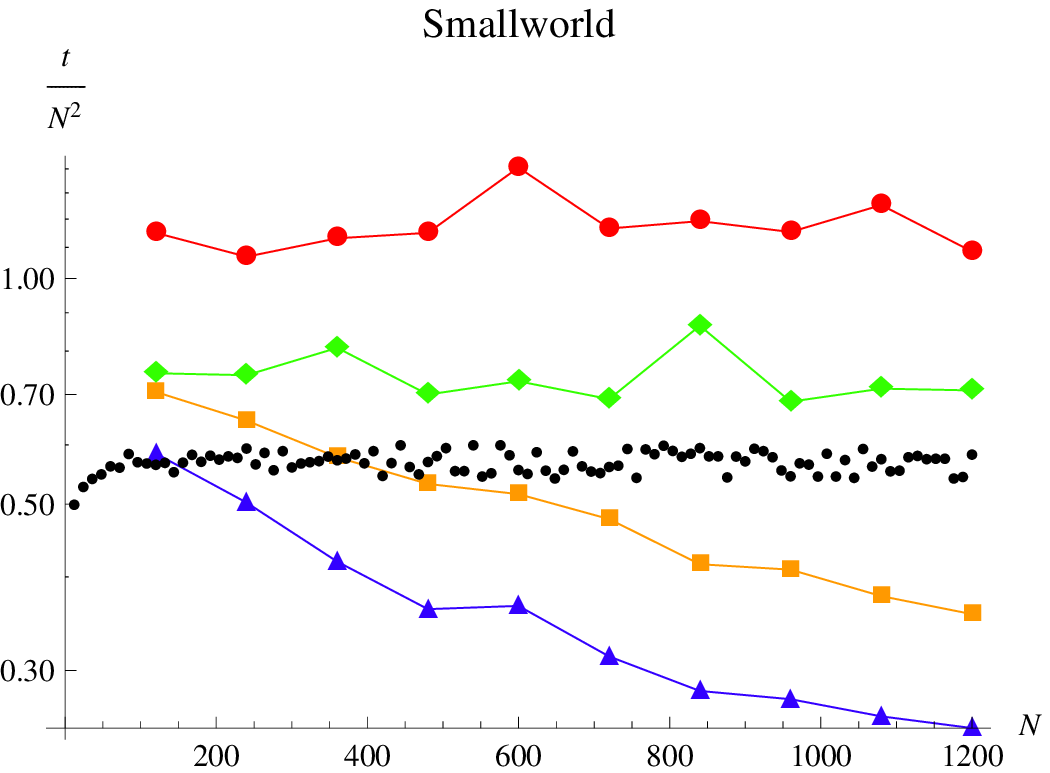,height=50mm}
\caption{(Color online.) Mean extinction times of the Moran process on networks. The plots are semi-logarithmic over $N$, and the MET is devided by $N^2$ for a better comparison with the result of well mixed populations and to better discern stabilizing effects. 
Red circles: $\overline{k}=4$, $\omega=0.05$, 
orange squares: $\overline{k}=4$, $\omega=0.40$, 
green rhombi: $\overline{k}=8$, $\omega=0.05$, 
blue triangles: $\overline{k}=8$, $\omega=0.40$, black points: well mixed population. Red circles on the upper edge of a plot in connection with a curve of a different colour mark are only marks that the curve is going on further, 
but it isn't depicted for a better presentation.
Averages are taken over a sample of 100 runs ($N>400$) 
or 1000 runs ($N=1000$).}
\label{Moran_Netzwerke}
\end{figure*}

From the results of our simulations (see Fig.~\ref{Moran_Netzwerke}), 
there are two 
immediate main observations: 
First, for some networks biodiversity is greatly stabilized as the MET increases exponentially with $N$, while for other networks biodiversity is only slightly stabilized or even destabilized. Second, the MET is no longer independent of the selection strength $\omega$ even in the zero-sum RPS case that we have
considered here. 
Stabilizing 
effects of the underlying structure are more distinct for greater $\omega$. 

Great stabilization of the biodiversity is only observed for networks with a 
heterogenous degree distribution and strong selection, while for networks 
with homogenous degree distribution we have no such strong stabilization. 
In fact, the networks with homogenous degree distribution can even lead to a slight destabilization of biodiversity for strong selection, as the MET grows slower than $N^2$, as we can see for the random regular graph. Even in the cases of the small world and the Gaussian network, which both have a small heterogeneity in their degree distribution, we have such a destabilization. Only for the square lattice we have a stabilization (although the lattice has a homogenous degree distribution, of course), 
but the stabilization is much weaker than in case of the random graph or the two scalefree networks. For the stabilization on the lattice, 
the large average shortest path 
length might play a role (which is proportional to $\sqrt{N}$ instead of $\log N$ as for random networks \cite{weglaenge zeigen}).

For small $\omega$, there is not a great change in the behaviour of the MET on all networks compared to unstructured populations
(Fig.~\ref{Moran_Netzwerke}).

Also, the correlation strength of the network degrees has only minor effect of the stabilization of biodiversity. 
The MET for an uncorrelated scalefree graph is, in all cases 
considered here, somewhat smaller than for the 
Barab\'asi-Albert scalefree graph, but in comparison to the other effects, 
mainly the enormous stabilization impact of scalefree networks, this is negligible.

\subsection{Local update process on networks}
\begin{figure*}%[pthb]
\epsfig{file=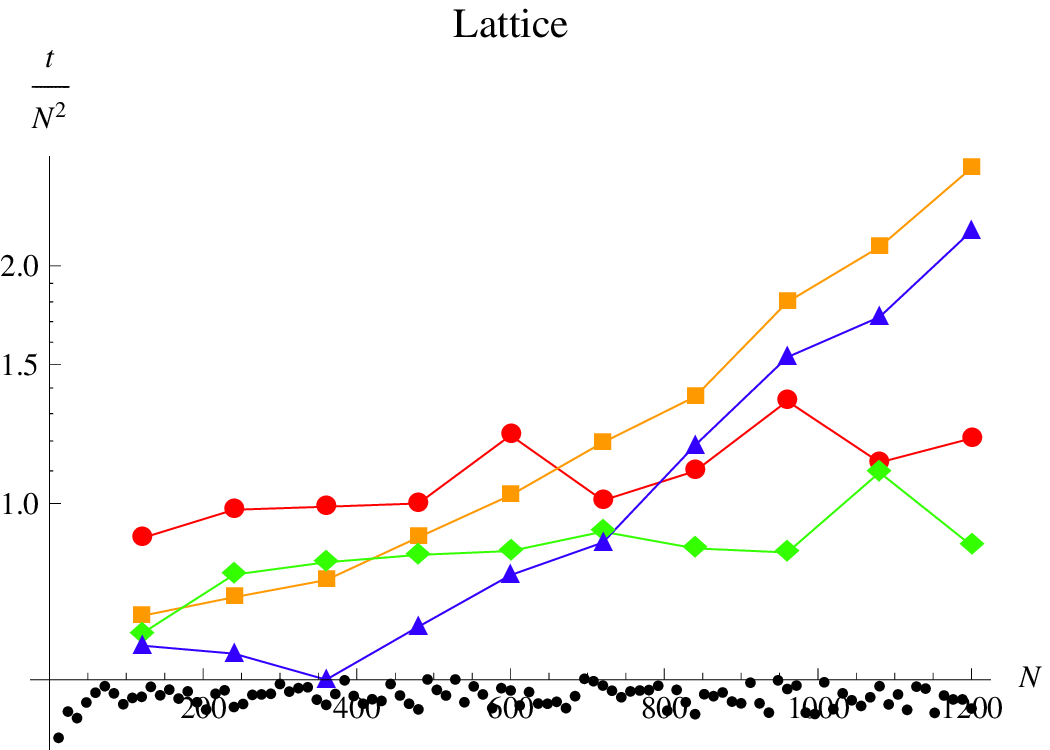,height=50mm}
\epsfig{file=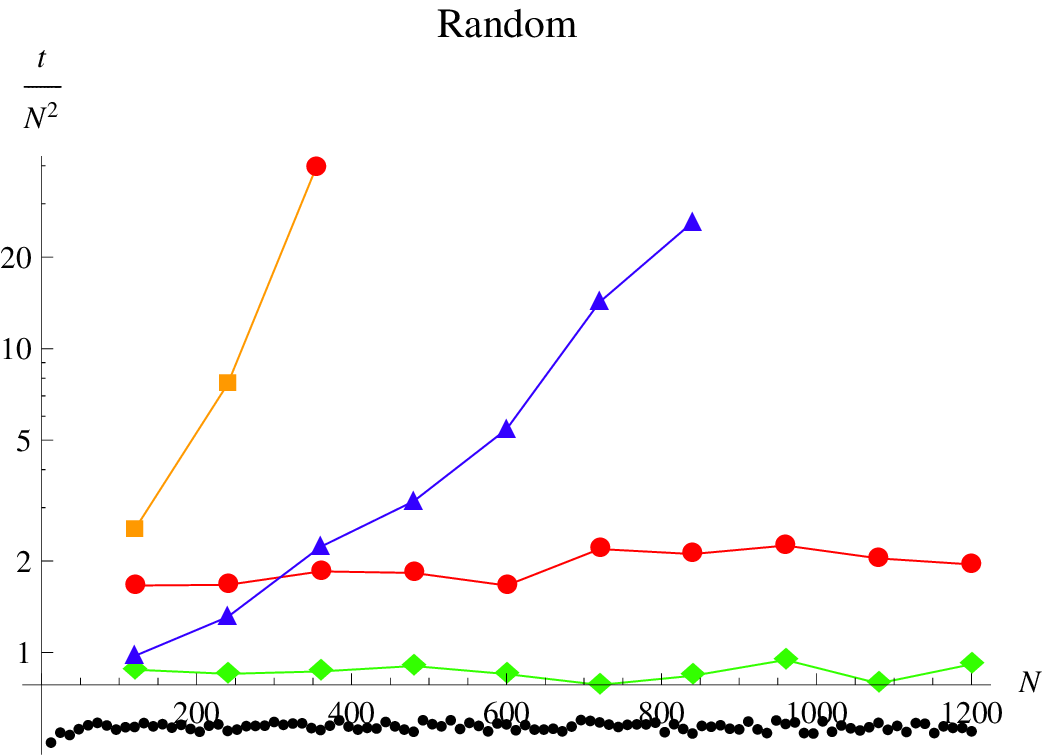,height=50mm}\\
\epsfig{file=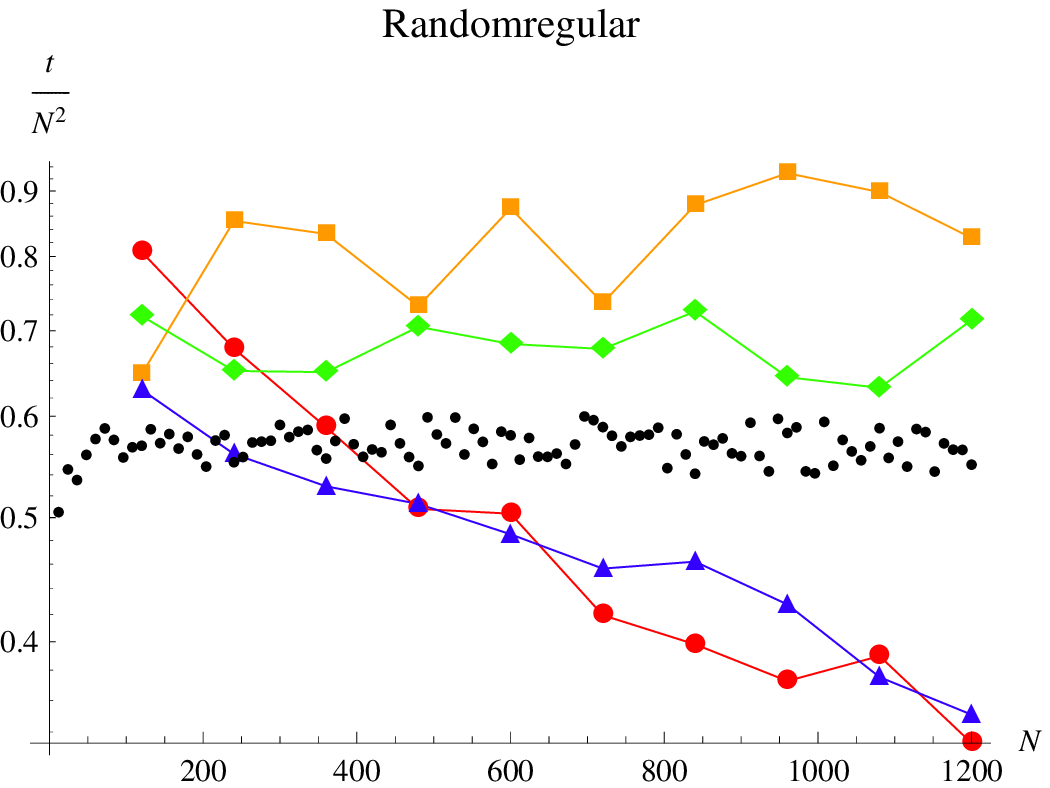,height=50mm}
\epsfig{file=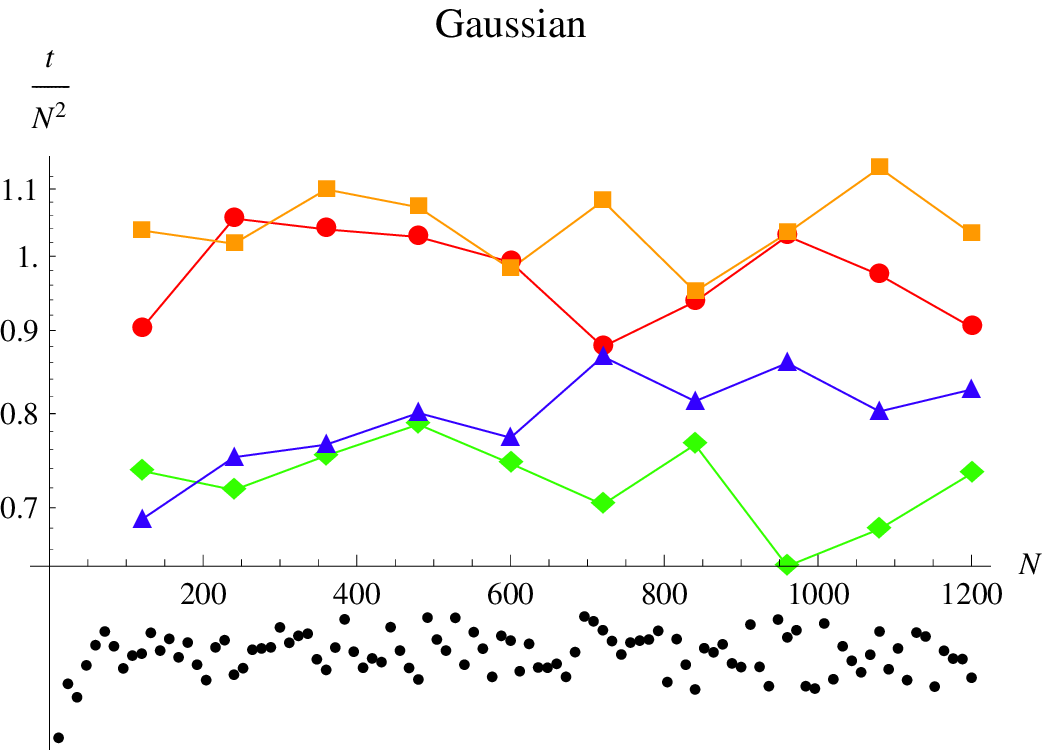,height=50mm}\\
\epsfig{file=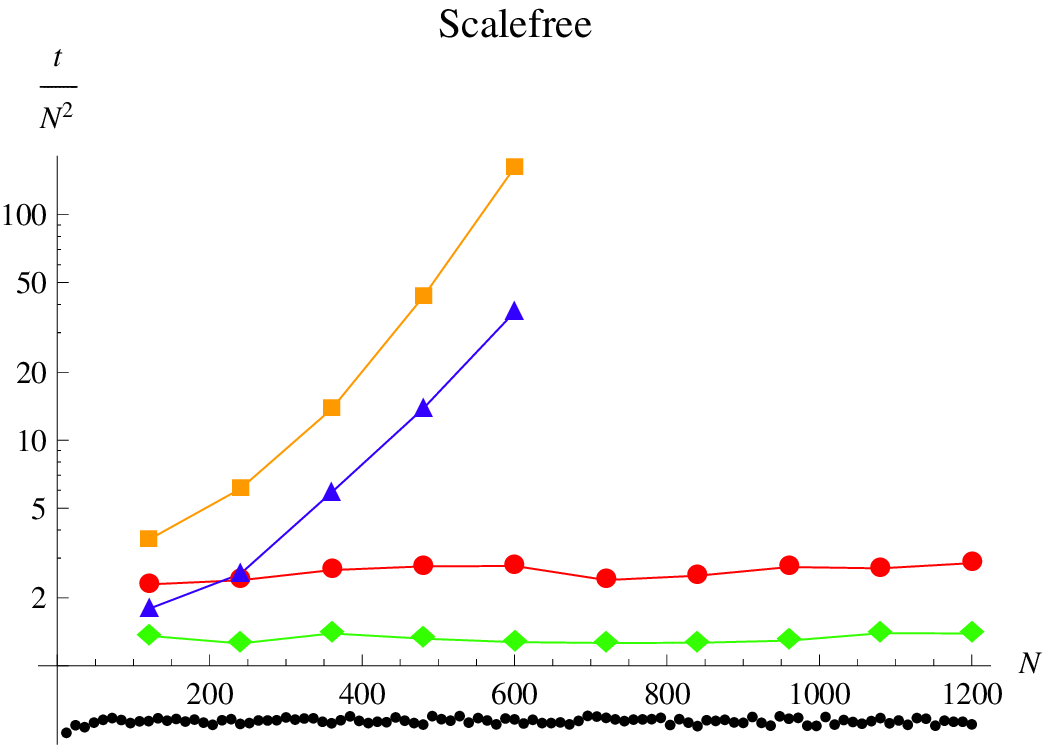,height=50mm}
\epsfig{file=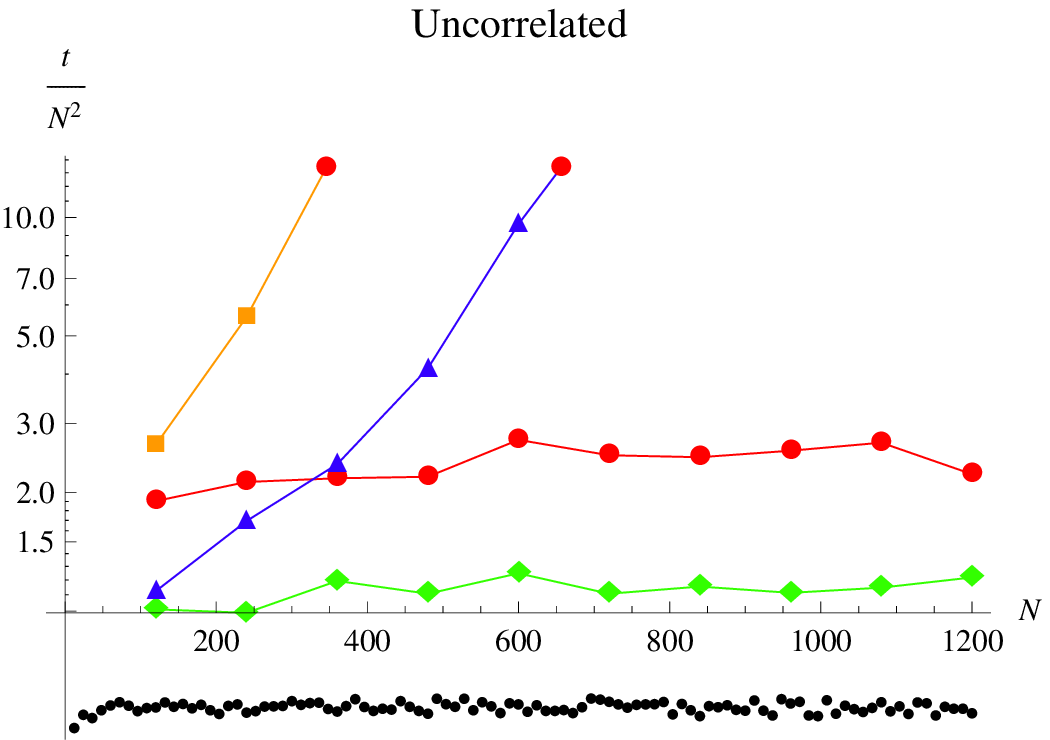,height=50mm}\\
\epsfig{file=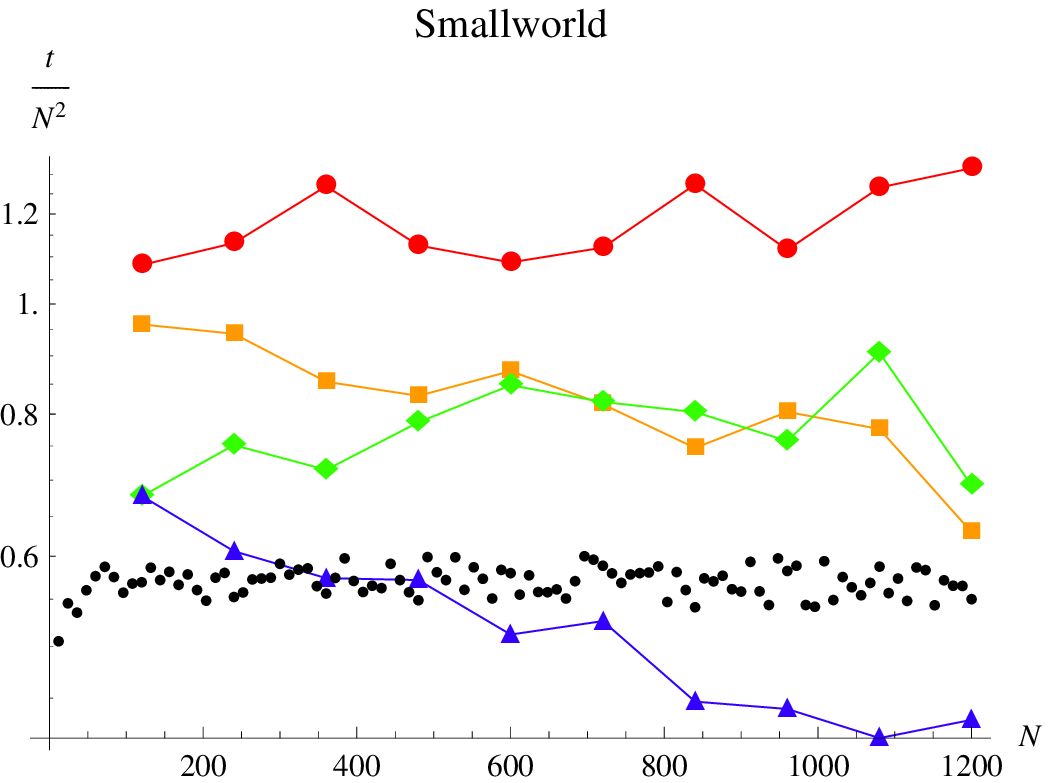,height=50mm}
\caption{(Color online.) Mean extinction times of the local update process on networks. The plots are semi-logarithmic over $N$, and the MET is devided by $N^2$ for a better comparison with the result of well mixed populations and to better discern stabilizing effects. 
Red circles:
$\overline{k}=4$, $\omega=0.05$, 
orange squares: $\overline{k}=4$, $\omega=0.50$, 
green rhombi: $\overline{k}=8$, $\omega=0.05$, 
blue triangles: $\overline{k}=8$, $\omega=0.50$, 
black points: well mixed population. Red circles on the upper edge of a plot in connection with a curve of a different colour marks that the curve is going on further, but it isn't depicted for a better presentation.}
\label{local_Netzwerke}
\end{figure*}
One of our main points of interest was whether the evolutionary process
itself influences the MET in the case of a networked population. 
The corresponding results for the Local Update are shown in 
Fig.~\ref{local_Netzwerke}.
There is no qualitative difference between the results in the 
Moran process and in the Local Update process. 
Same as for the Moran process, 
heterogenous degree distributions cause an enormous 
stabilization of the biodiversity for strong selection 
(the MET grows exponentially with $N$), while homogenous degree 
distributions do not. The MET is now dependent on the selection 
strength as well, and for small $\omega$ it is proportional to $N^2$ as in the neutral case. The only difference is that stabilizing effects are not as strong as in the Moran process for the same or even a 
bit larger $\omega$.

\subsection{SPD-process on networks}
\begin{figure*}%[pthb]
\epsfig{file=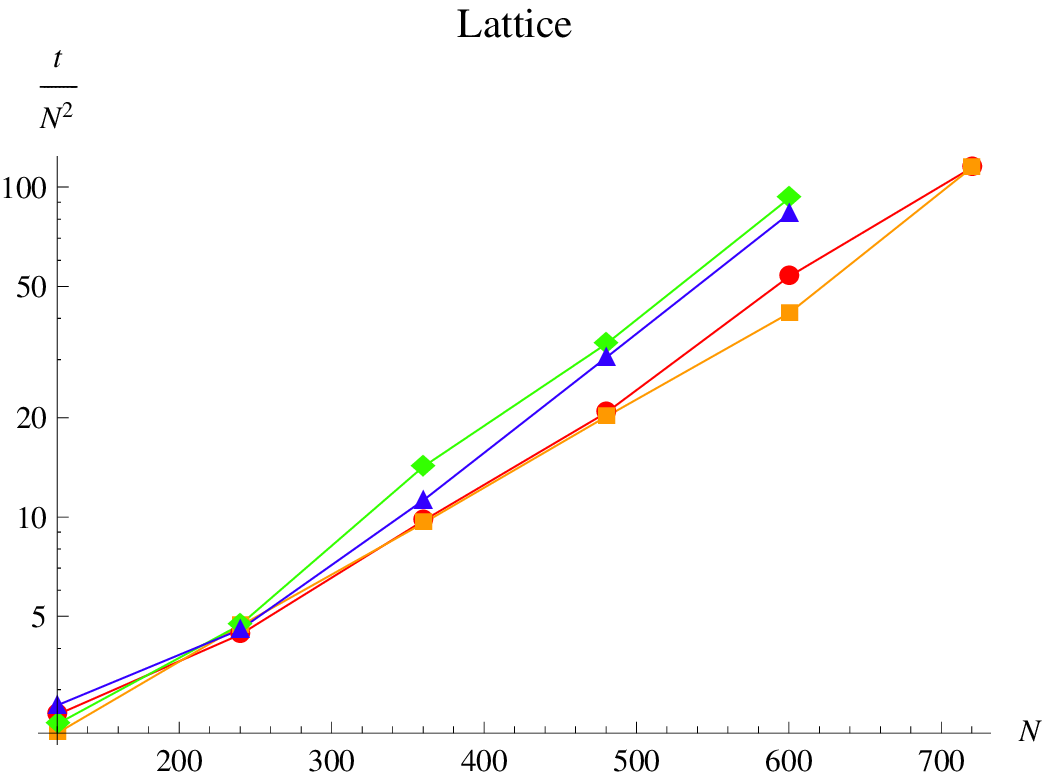,height=50mm}
\epsfig{file=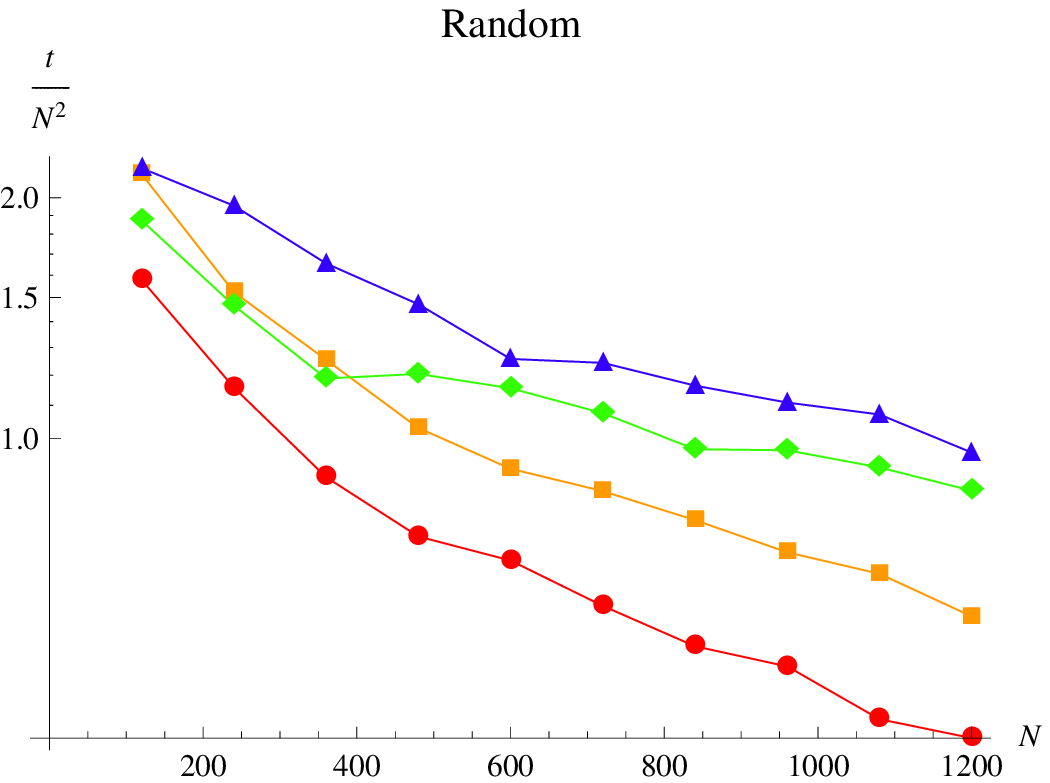,height=50mm}\\
\epsfig{file=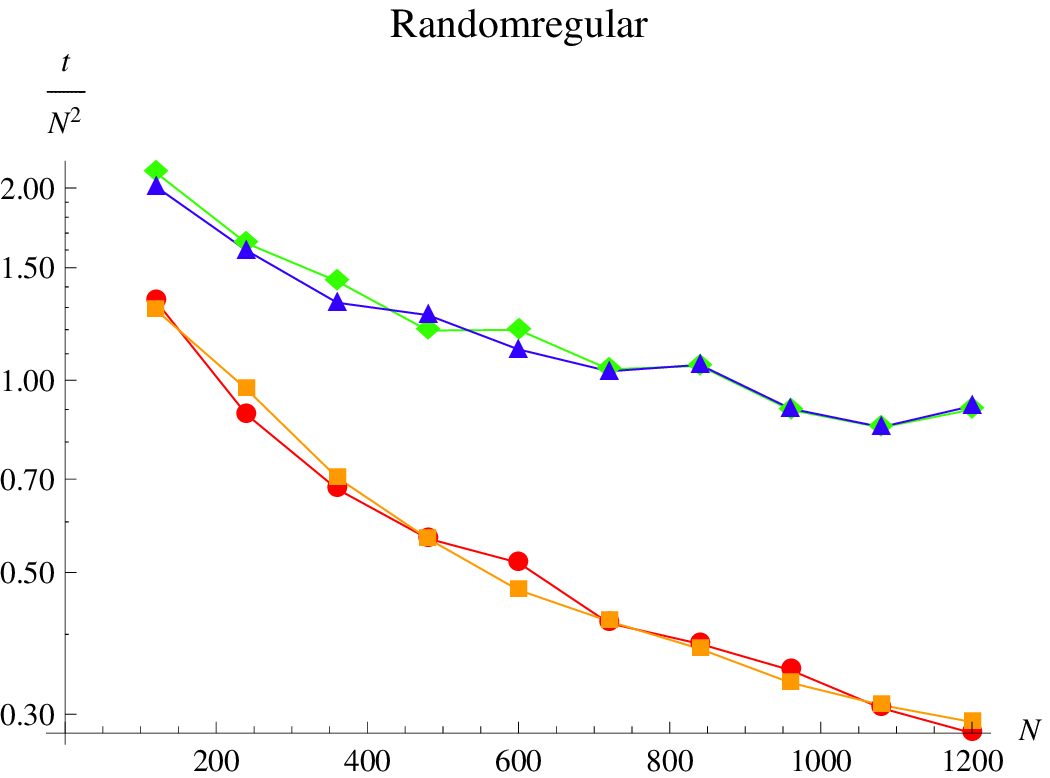,height=50mm}
\epsfig{file=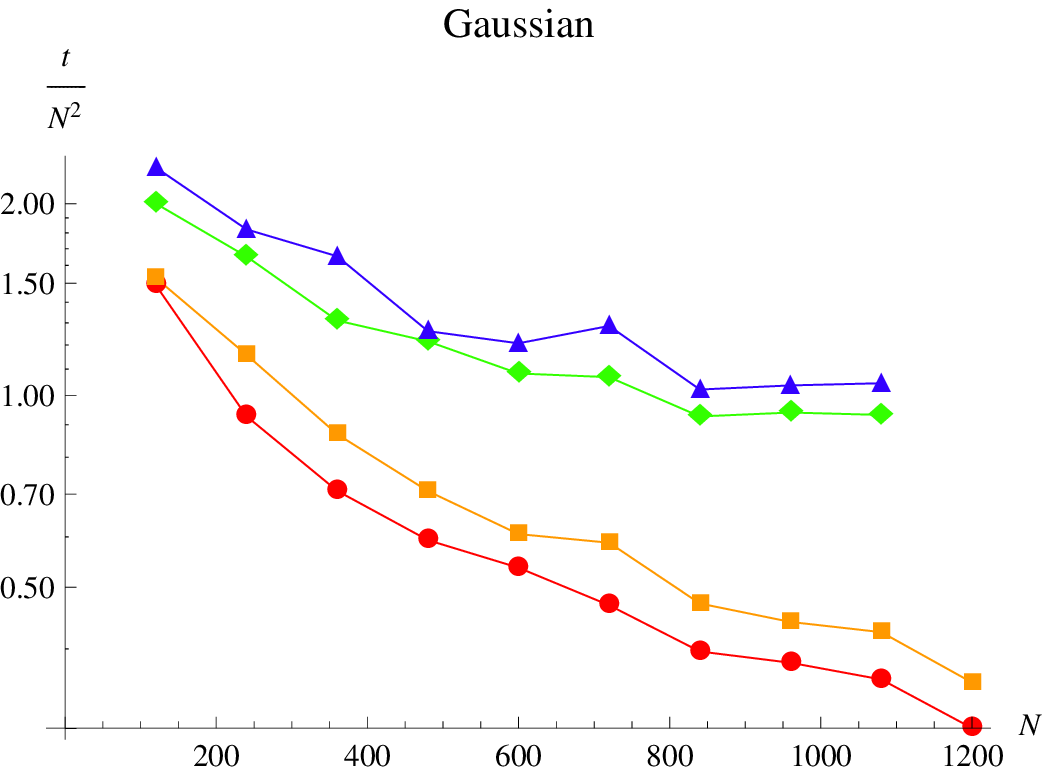,height=50mm}\\
\epsfig{file=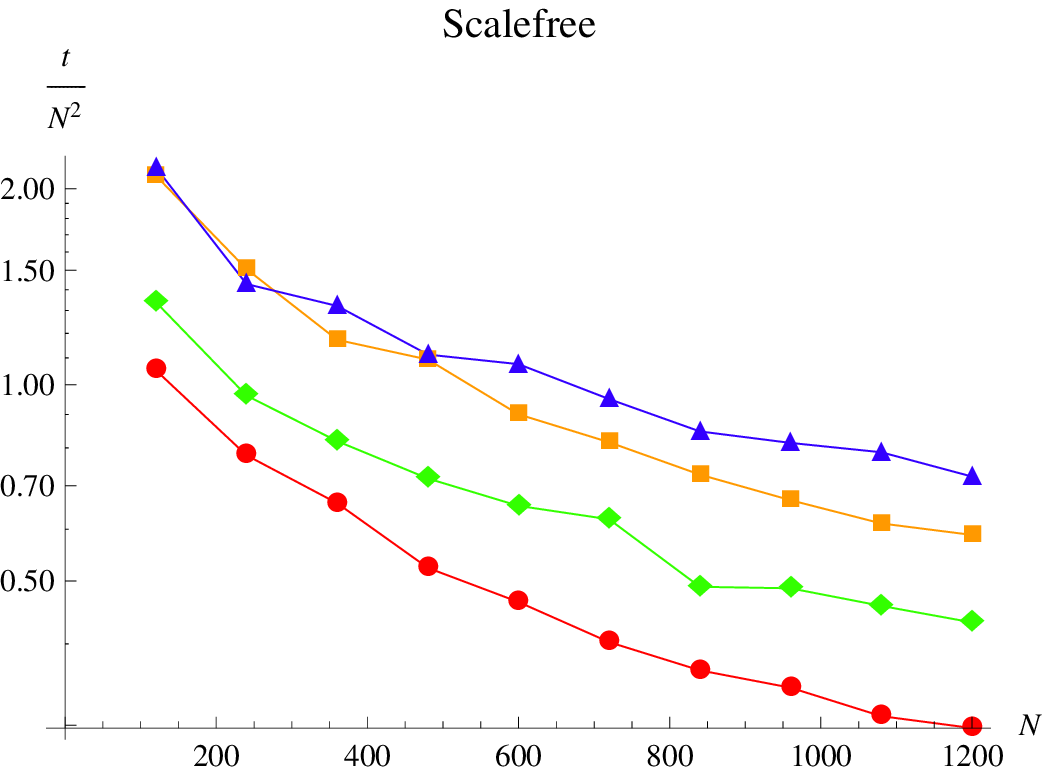,height=50mm}
\epsfig{file=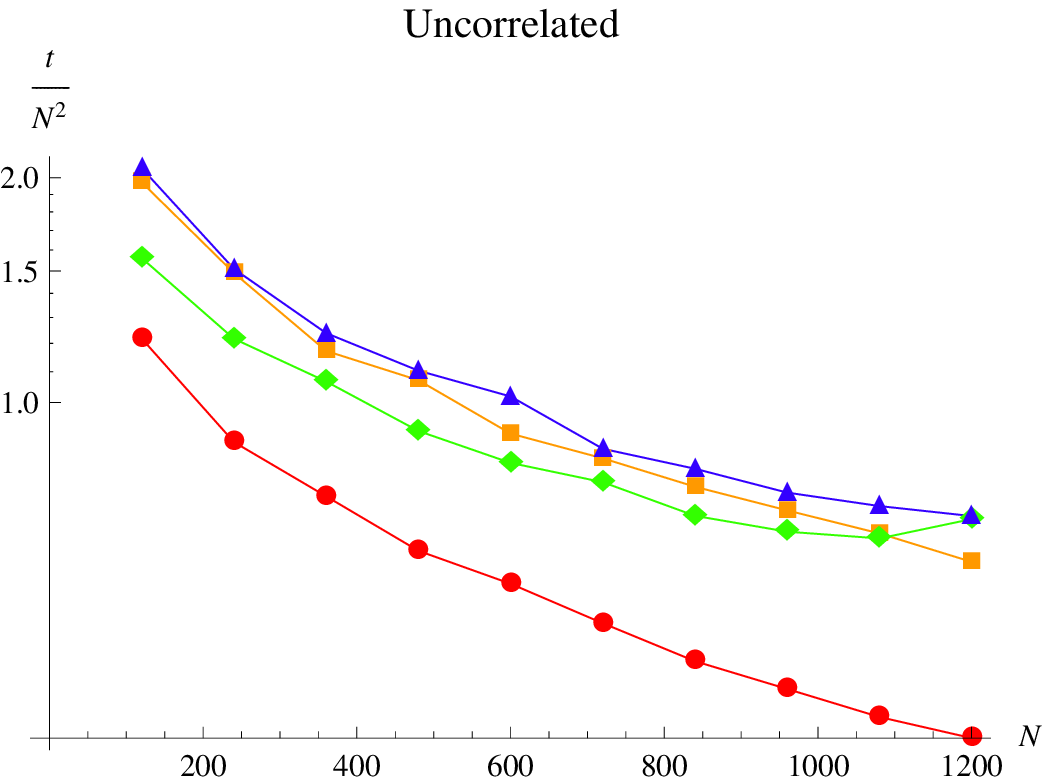,height=50mm}\\
\epsfig{file=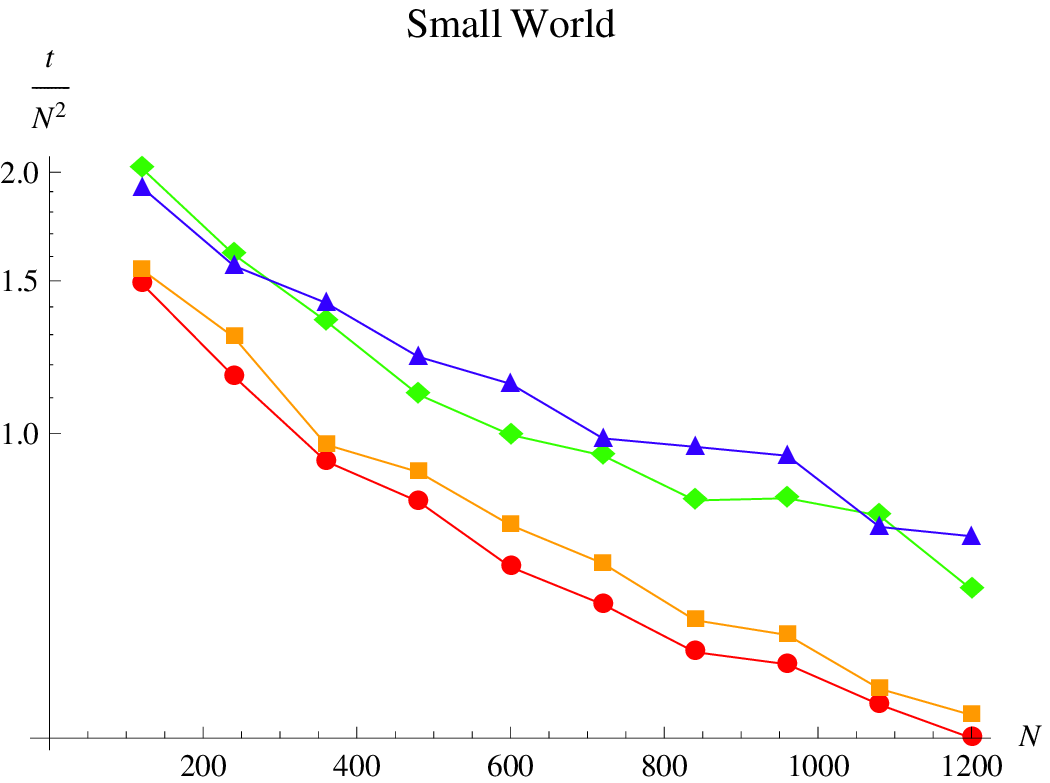,height=50mm}
\caption{(Color online.) Mean extinction times of the SPD-process on networks. The plots are semi-logarithmic over $N$, and the MET is devided by $N^2$ for a better comparison with the result of the other processes. 
Red circles: $\overline{k}=4$, $a=0.0$, 
orange squares: $\overline{k}=4$, $a=1.0$, 
green rhombi: $\overline{k}=8$, $a=0.0$, 
blue triangles: $\overline{k}=8$, $a=1.0$.}
\label{Szolnoki_Netzwerke}
\end{figure*}
The SPD-process behaves completely different,
as shown in Fig.\ \ref{Szolnoki_Netzwerke}.
We recall  that for this update rule the 
quite general observation that spatial discretization 
can lead to a stabilization of biodiversity does not hold. 
Even 
ignoring this and taking  a MET scaling for large $N$ 
like 
the $N^2$ 
scaling from Moran process and Local Update
as reference, 
for all studied graphs except the lattice we have a destabilization, 
and the MET grows slower than $N^2$ with only small differences between 
the different networks. Only in the lattice case the MET grows exponentially 
with $N$.

Changing the parameter $a$ has minor influence on the properties of the MET than the selection strength $\omega$ in the Moran and the 
Local Update process. For degree homogenous networks, 
the MET is even 
independent of $a$. For networks with a strong heterogeneity in the degree distribution the MET grows with $a$ for the same value of $N$.

There is also some influence of the correlation of network degrees. For the uncorrelated scalefree network, the influence of the parameter $a$ on the MET is not as great as for the 
Barab\'asi-Albert network, and the METs for the same $N$, but different $a$ and $\overline{h}$ are closer together. But compared to the other effects, this is again negligible.

\section{Theoretical approach}
\subsection{Moran process}
\subsubsection{General case}
For a version of the RPS game with constant reaction rates it has already been shown that networks with heterogenous degree distribution effect a stabilization of biodiversity \cite{Netzwerk Stabilitaet}. This corresponds with our simulation results so we have adapted the approach from 
\cite{Netzwerk Stabilitaet} to underpin the shown behaviour theoretically.

First let us define the probability density of the strategy $\alpha$ on all 
nodes with degree $k$ as $\rho_{\alpha,k}, 
where \alpha=(0,1,2)$. Here we again use the notation $\alpha$ for an arbitrary strategy, 
$\beta$ for the strategy 
dominated by $\alpha$, and $\gamma$ for the dominating strategy. Then the probability $\theta_{\alpha^{\prime}}$ that a neighbored node of an arbitrary vertex plays strategy $\alpha^{\prime}$ reads
\begin{equation}
\theta_{\alpha^{\prime}}=\sum_{k^{\prime}} k^{\prime} p_{k^{\prime}} \rho_{\alpha^{\prime}}/\langle k\rangle.
\end {equation}
Here $p_k$ is the probability of finding a vertex with degree $k$. Dividing by the average vertex degree, $\langle k\rangle=\sum_k k p_k$, assures the standardization. 
Further, by the conservation of the total density one of the quantities $\rho_{\alpha,k}$ and of $\theta_\alpha$ is given so that we can eliminate $\rho_{2,k}$ and $\theta_2$ using $\rho_{0,k}$ and $\rho_{1,k}$ as variables, or $\theta_0$ and $\theta_1$, respectively
\begin{eqnarray}
\rho_{2,k}&=&1-\rho_{0,k}-\rho_{1,k}\\
\theta_2&=&1-\theta_0-\theta_1
\end{eqnarray}
The average payoff of a player of strategy $\alpha$ on a node with degree $k$ is then given by
\begin{equation}
\pi_{\alpha,k} =\theta_\beta-\theta_\gamma
\end{equation}
We find that the average payoff is independent of the vertex degree $k$ 
so we can drop the index $k$, $\pi_{\alpha,k}\equiv \pi_\alpha$. 
With this we can compute the average payoff of neighbours of a vertex
\begin{eqnarray}
\langle\pi\rangle&=&\frac{\sum_{\alpha=0}^2\sum_{k^{\prime}}\pi_\alpha\frac{p_{k^{\prime}}k^{\prime}}{\langle k\rangle}\rho_{\alpha,k^{\prime}}k}{k}\nonumber\\
&=&\sum_{\alpha=0}^2\sum_{k^{\prime}}\pi_\alpha\frac{p_{k^{\prime}}k^{\prime}}{\langle k\rangle}\rho_{\alpha,k^{\prime}}\nonumber\\
&=&\sum_{\alpha=0}^2\left(\theta_\beta-
\theta_\gamma\right)\sum_{k^{\prime}} k^{\prime} p_{k^{\prime}} \rho_{\alpha^{\prime}}/\langle k\rangle\nonumber\\
&=&\sum_{\alpha=0}^2\left(\theta_\beta-
\theta_\gamma\right)\theta_\alpha.
\end{eqnarray}
By using $k^{\prime} p_{k^{\prime}} \rho_{\alpha^{\prime}}/\langle k\rangle$ instead of $p_{k^{\prime}}$ we consider that is more likely to have a vertex with a high degree as a neighbour than one with a smaller degree if both are equally frequent. Thus we find the probability for a player with strategy $\alpha$ sitting on a node with degree $k$ to 
reproduce as
\begin{equation}
w_{\alpha,k}=\frac{1}{2}p_k\rho_{\alpha,k}\frac{1-\omega+\omega \pi_\alpha}{1-\omega+\omega\langle\pi\rangle},
\end{equation}
and the rate with that a player of strategy $\alpha$ sitting on a vertex with degree $k$ conveys his strategy to a node with degree $k^{\prime}$, which has played strategy $\alpha^{\prime}$ before, reads
\begin{eqnarray}
T_{k,k^{\prime}}^{\alpha^{\prime}\rightarrow\alpha}&=&\frac{p_{k^{\prime}}\rho_{\alpha^{\prime},k^{\prime}}}{\langle k\rangle}w_{\alpha,k}\\
&=&\frac{1}{2} ~
\frac{p_{k^{\prime}}\rho_{\alpha^{\prime},k^{\prime}}}{\langle k \rangle}p_k\rho_{\alpha,k}
\frac{1-\omega+\omega \pi_\alpha}{1-\omega+\omega\langle\pi\rangle}\nonumber
\end{eqnarray}
For $p_k=\delta_{k,k_0}$ ($\delta_{k,k_0}$ is the Kronecker symbol) this reduces to
\begin{equation}
\frac{1}{2}
T^{\alpha^{\prime}\rightarrow\alpha}=\rho_{\alpha}\rho_{\alpha^{\prime}}
\frac{1-\omega+\omega\pi_\alpha}{1-\omega+\omega\langle\pi\rangle}.
\end{equation}
Here we have omitted any $k$-indices because there is only one degree left. The payoffs reduce to 
$\pi_\alpha=\rho_\beta-\rho_\gamma$
and $\langle\pi\rangle=\sum_{\alpha=0}^2 \pi_\alpha\rho_\alpha$,
and one obeys the hopping rates for the Moran process in well mixed populations \cite{CT08} independent of $k_0$. This rate describes how likely it is that a player with strategy $\alpha^{\prime}$ is replaced by a copy of another player with strategy $\alpha$ in a single time step.

Let us now consider a heterogenous degree distribution. 
In the simplest case, such a distribution consists of only two degrees $k_1$ and $k_2$ with frequencies $p_1=:p$ and $p_2=1-p_1=1-p$, respectively. 
The replicator equation that we obtain from the master equation
in the limit $N\rightarrow\infty$ for the changing of the frequency of the strategy $0$ on nodes with degree $k_1$ 
reads (for the methodology see e.g.\
\cite{vankampen,TCH05,mckane05}):
\begin{eqnarray}
\dot\rho_{0,k_1}&=&T_{k_1,k_1}^{1\rightarrow 0}+T_{k_1,k_1}^{2\rightarrow 0}-T_{k_1,k_1}^{0\rightarrow 1}-T_{k_1,k_1}^{0\rightarrow 2}\\
&&+T_{k_2,k_1}^{1\rightarrow 0}+T_{k_2,k_1}^{2\rightarrow 0}-T_{k_2,k_1}^{0\rightarrow 1}-T_{k_2,k_1}^{0\rightarrow
2}.\nonumber
\end{eqnarray}
Inserting our special degree distribution, after 
a longish but straightforward calculation we find
\begin{eqnarray}
\dot\rho_{0,k_1}&=&\rho_{0,k_1}\frac{k_1 p^2}{2\langle k\rangle(\Gamma+\langle\pi\rangle)}(\pi_0-\langle\pi\rangle_1)\\
&&+\frac{k_1 p(1-p)}{2\langle k\rangle(\Gamma+\langle\pi\rangle)}
\left(\Gamma\left(\rho_{0,k_2}-\rho_{0,k_1}\right)\right.\nonumber\\
&&\left.+\rho_{0,k_2}\pi_0-\rho_{0,k_1}\langle\pi\rangle_2\right).\nonumber
\end{eqnarray}
Here $\Gamma=\frac{1-\omega}{\omega}$ is the background fitness. $\langle\pi\rangle_1=\pi_0\rho_{0,k_1}+\pi_1\rho_{1,k_1}+\pi_2\rho_{2,k_1}$ is the mean fitness of all players on vertices with degree $k_1$, and $\langle\pi\rangle_2$ correspondingly for $k_2$. 
Then one can compute the other equations in an analogical way 
or by cyclically permuting 
the indices. 
For each vertex degree, 
one of the equations of motion
needs not to be taken into account by the constraint $\rho_{0,k_i}+\rho_{1,k_i}+\rho_{2,k_i}=1$. For $k_1=k_2$ or alternatively $p=1$ this flattens to the well known adjusted replicator equation in the case of well mixed populations, 
as expected \cite{TCH05,TCH06}. 
Likewise, the fixed point of this replicator equation at $\left(\frac{1}{3},\frac{1}{3},\frac{1}{3},\frac{1}{3}\right)$ 
does not alter, we now just have two densities for every strategy. 
But now the velocity of the reaction is directly dependent of $\omega$ because it is no longer 
possible to absorbe the strength of selection by a dynamical rescaling of time, 
as it is possible for well mixed populations.

But what about the stability of this inner fixed point? Within a linear stability analysis, let us linearize the replicator equation around its fixed point
\begin{equation}
\dot{\vec{\rho}}=\underline{\underline{A}}\vec{\rho},
\end{equation}
with $\vec{\rho}=\left(\rho_{0,k_1},\rho_{1,k_1},\rho_{0,k_2},\rho_{1,k_2}\right)$ and $\underline{\underline{A}}$ being the Jacobian of the system with $A_{ij}=\frac{\partial \rho_i}{\partial\rho_j}\vert_{\vec{\rho}=\left(\frac{1}{3},\frac{1}{3},\frac{1}{3},\frac{1}{3}\right)}$. In our case, we find $\underline{\underline{A}}$ as
\small
\begin{equation}
\left(
\begin{array}{cccc}
 \frac{k_1 (p-1) p}{2 \langle k\rangle} &
   \frac{k_1^2 p^2}{6 \langle k\rangle^2 \Gamma } &
   -\frac{k_1 (p-1) p}{2 \langle k\rangle} &
   -\frac{k_1 k_2 (p-1) p}{6 \langle k\rangle^2
   \Gamma } \\
 -\frac{k_1^2 p^2}{6 \langle k\rangle^2 \Gamma } &
   \frac{k_1 (p-1) p}{2 \langle k\rangle} &
   \frac{k_1 k_2 (p-1) p}{6 \langle k\rangle^2
   \Gamma } & -\frac{k_1 (p-1) p}{2 \langle k\rangle}
   \\
 -\frac{k_2 (p-1) p}{2 \langle k\rangle} &
   -\frac{k_1 k_2 (p-1) p}{6 \langle k\rangle^2
   \Gamma } & \frac{k_2 (p-1) p}{2 \langle k\rangle} &
   \frac{k_2^2 (p-1)^2}{6 \langle k\rangle^2 \Gamma }
   \\
 \frac{k_1 k_2 (p-1) p}{6 \langle k\rangle^2
   \Gamma } & -\frac{k_2 (p-1) p}{2 \langle k\rangle}
   & -\frac{k_2^2 (p-1)^2}{6 \langle k\rangle^2 \Gamma
   } & \frac{k_2 (p-1) p}{2 \langle k\rangle}
\end{array}\right).
\label{Stablitat_Moran_Matrix}
\end{equation}
\normalsize
To reason about the stability we need to know the signs of the real parts of the 
eigenvalues $\lambda_i$ of this matrix, but for these eigenvalues we derive 
quite complicated terms, the signs of which cannot be easily designated (but see Appendix \ref{Eigenwerte_Moran}). 
The Routh-Hurwitz criterion as an alternative solution 
does not
simplyify the situation considerably.

Approximating an upper bound of the real parts of all eigenvalues, it is possible to compute a region in the parameter space where the inner fixed point is guaranteed to be asymptotically stable, namely, for
\begin{equation}
0.5\leq p<\frac{k_{max}+k_{min}}{k_{max}+k_{min}+3\frac{k_{max}^2}{\langle k\rangle}}.
\end{equation}
For a given value of $p$ this determines the ratio of $k_{max}$ and $k_{min}$, 
or, the other way round, if we want to build up a network with chosen values 
for $k_{max}$ and $k_{min}$, this inequality tells us what frequencies of 
both vertex degrees are necessary to stabilize biodiversity. 
So, for example, if we choose $k_1=4$ and both vertex degrees 
should be equally frequent, $k_2$ must be at least $7$ to fulfill 
this inequality. If a higher frequency of $k_1$ is desired, $k_2$ 
must even be larger.

But as this is just a rough approximation, by inserting numerical values 
we have found that almost all eigenvalues are $<0$, except for the trivial 
cases $k_1=k_2$, $p=0$ or $p=1$. So for all values except these mentioned 
cases, we have found the fixed point to be asymptotically stable. 
Here, we have found numerically that the real parts of the eigenvalues 
become more negative as 
the difference of $k_1$ and $k_2$ grows in absolute size, and both vertex degrees
become equally likely to be found in the network. This explains as well 
why the stabilization of biodiversity is greater in the case of the scalefree 
graphs than for the ER-random graph, or it is in both cases greater than for 
the network with a gaussian degree distribution, which has only as small 
heterogeneity in its degree distribution.

\subsubsection{Small selection limit}
As we have seen in the previous section, for small $\omega$ the MET is almost proportional to $N^2$ as in the well mixed case, 
and depends only marginally on the underlying network structure. 
We can intuitively understand this as follows: 
If $\omega$ is small, than every transition probability 
is $\approx\frac{1}{2}$, and the corresponding replicator 
equation in our system reads
\begin{equation}
\dot\rho_{\alpha,k}=0
\end{equation}
for all values of $\alpha$ and $k$. 
Hence, each point of the state space is a neutrally stable fixed point 
in the limit $N\to\infty$. 
Therefore there should be no stabilization 
impact compared to the well mixed case, 
as there each point in the simplex belongs 
to a neutrally stable limit cycle, so in both cases a perturbation 
is neither expected to decay nor to grow in time.

\subsection{Local update process}
\begin{figure}[htbp]
\epsfig{file=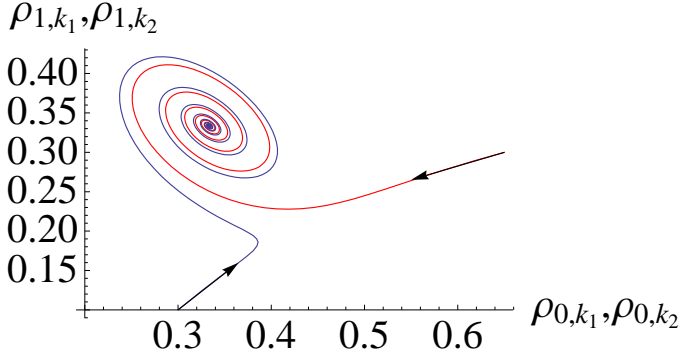,width=0.5\textwidth}
\caption{(Color online.) A plot of a numerical solution of the replicator 
equation for the local update process on a network with only two different 
vertex degrees 
(eq.\ \ref{eqB2})
with $p=0.5$, $k_1=4$, $k_2=8$,
$\omega=0.4$, $\rho_{0,k_1}(0)=0.65$, $\rho_{1,k_1}(0)=0.3$, $\rho_{0,k_2}(0)=0.3$, $\rho_{1,k_2}(0)=0.1$, red for the density of strategy $0$ on nodes with degree $k_1$, black for the corresponding value on vertices with degree $k_2$}
\end{figure}
Following the same scheme as for the Moran process, one can derive similar 
results (see Appendix \ref{local_theo} for details). As in the Moran process, 
a degree heterogeneity leads to a stabilization of biodiversity, but again the 
eigenvalues of the Jacobian are too byzantine for catching the signs 
of the real parts easily.

\subsection{SPD-process}

For this update process, we will 
use a simpler method 
to achieve a theoretical understanding of some of the numerical results. In the case of both chosen vertices having the same degree, the probability for the node $i$ to change its strategy to the strategy of $j$ simplifies to
\begin{equation}
W_{S(i)\rightarrow S(j)}\vert_{h(i)=h(j)}=\frac{\pi_j-\pi_i}{
2
k_{max}
}.
\label{eq29}
\end{equation}
So for the lattice and the random regular graph, where all vertex degrees are the same, any dependence of $a$ drops out. Note that in this equation and in this whole subsection, when talking about $\pi_i$ we consider the not modified payoffs. For $a=1$ one achieves the same term as in the case of 
degree homogenous networks
(eq.~\ref{eq29}), 
independent of the network, while for $a=0$ we find
\begin{equation}
W_{S(i)\rightarrow S(j)}\vert_{a=0}=\frac{\frac{\pi_j}{k(j)}-
\frac{\pi_i}{k_{max}}}{
2}
\end{equation}
or
\begin{equation}
W_{S(i)\rightarrow S(j)}\vert_{a=0}=\frac{\frac{pi_j}{k_{max}}-
\frac{\pi_i}{k(i)}}{
2},
\end{equation}
respectively, depending on which of the vertex degrees is greater. If we compare the probabilities for $a=1$ and for $a=0$ by 
computing the ratio of both, we can 
analyze the impact of the parameter $a$ and the underlying network structure on the MET. We find
\begin{equation}
\frac{W_{S(i)\rightarrow S(j)}\vert_{a=1}}{W_{S(i)\rightarrow S(j)}\vert_{a=0}}=\frac{k(j)(\pi_i-\pi_j)}{k(j)\pi_i-k_{max}\pi_j}
\end{equation}
or
\begin{equation}
\frac{W_{S(i)\rightarrow S(j)}\vert_{a=1}}{W_{S(i)\rightarrow S(j)}\vert_{a=0}}=\frac{k(j)(\pi_i-\pi_j)}{k_{max}\pi_i-k(i)\pi_j},
\end{equation}
respectively. If we account for the fact that $\pi_j>\pi_i$ must hold, then the first fraction is $<1$ for $\pi_j>0$ and $>1$ for $\pi_j<0$. 
Hence, if aside from $a$ all conditions are the same, for $a=1$ it is 
more likely that a strategy with a lower payoff reproduces, 
if $k(j)<k(i)$) than for $a=0$. 
On the other hand under these conditions for $a=1$ it is less likely that a strategy with a high payoff reproduces, than for $a=0$. 
In the second case, the fraction is $<1$ for $\pi_j>0$ and $\pi_i<0$ and $>1$ for all other cases. 
Hence for $a=1$ it is more likely that a strategy with a high payoff 
replaces a strategy with a smaller payoff from a node with a smaller 
vertex degree, than for $a=0$. 
And for $a=1$ under this conditions it is as much more likely that a strategy with a smaller payoff reproduces. 
Taken together, for $a=1$ strategies with a smaller payoff have a bigger chance to reproduce, and they are 
less likely to be replaced by a strategy with a higher payoff, than for $a=0$. 
Hence on degree-heterogenous networks the MET should be greater for $a=1$
than for $a=0$, as it is confirmed by the simulation results 
(Fig.\ \ref{Szolnoki_Netzwerke}).

\section{Conclusions}
Cyclic coevolutionary dynamics manifests an interesting class
of dynamical systems that are discussed as models 
to explain long-term stabilization of coexistence
of species or strategies.
In this paper we have shown that, for the coevolutionary RPS game on complex
networks,
there are considerable differences in the 
mean extinction time (MET) between different networks, 
and for different processes as well. 
The simple statement ``Spatial discretization stabilizes biodiversity'' 
must be refined in a more sophisticated way. 
While for example in the Moran and in the 
local update process degree-heterogenous 
networks preserve biodiversity 
(the MET grows exponentially with the number $N$ of players), 
in the SPD-process they do not, and even degree-homogeneous 
networks may destabilize while the lattice still stabilizes.
In summary, we have shown that, for the cyclic evolutionary RPS on networks,  
there is a striking influence on the mean extinction time characteristics of 
the spatial structure \textit{and} the update rule as well. 

\appendix
\section{\label{Eigenwerte_Moran}Eigenvalues of the Jacobian in equation
(\ref{Stablitat_Moran_Matrix})}

With the help of (e.g.) Mathematica\texttrademark{} 
%\cite{Mathematica} 
one can easily compute the eigenvalues $\lambda_i$ of the Jacobian in 
equation (\ref{Stablitat_Moran_Matrix}):
\begin{flalign}
\lambda_1=&-\frac{1}{12 \langle k\rangle^2 \Gamma }\left(i \left(k_2^2 (p-1)^2+k_1^2 p^2\right)\right.&\\
&\left.-3
   (k_1+k_2) \langle k\rangle (p-1) p \Gamma
+\sqrt{c_1-i c_2}\right)\nonumber&\\
\lambda_2=&\frac{1}{12 \langle k\rangle^2 \Gamma }\left(-i
   \left(k_2^2 (p-1)^2+k_1^2 p^2\right)\right.&\\
&\left.+3
   (k_1+k_2) \langle k\rangle (p-1) p \Gamma+\sqrt{c_1-i c_2}\right)&\nonumber\\
\lambda_3=&\frac{1}{12 \langle k\rangle^2 \Gamma }\left(i
   k_2^2 (p-1)^2+3 k_2 \langle k\rangle p \Gamma 
   (p-1)\right.&\\
&\left.+k_1 p (i k_1 p+3 \langle k\rangle (p-1) \Gamma
   )-\sqrt{c_1+i c_2}\right)&\nonumber
\end{flalign}
\begin{flalign}
\lambda_4=&\frac{1}{12 \langle k\rangle^2 \Gamma }\left(i
   k_2^2 (p-1)^2+3 k_2 \langle k\rangle p \Gamma 
   (p-1)\right.&\\
&\left.+k_1 p (i k_1 p+3 \langle k\rangle (p-1) \Gamma
   )+\sqrt{c_1+i c_2}\right)&\nonumber
\end{flalign}
with the abbreviations
\begin{eqnarray}
c_1&=&-\left(k_2^2 (p-1)^2+k_1^2 p^2\right)^2\\
&&+9
   (k_1+k_2)^2 \langle k\rangle^2 (p-1)^2 p^2 \Gamma ^2\nonumber\\
c_2&=&6 \langle k\rangle (p-1) p \left(p^2 k_1^3+k_2 (p-2) p
   k_1^2\right.\\
&&\left.+k_2^2 \left(p^2-1\right) k_1+k_2^3
   (p-1)^2\right) \Gamma.\nonumber
\end{eqnarray}
The real parts $\Re\lambda_i$ of this eigenvalues are then given by
\begin{eqnarray}
\Re{\lambda_1}&=&\frac{k_1 p^2}{4 \langle k\rangle}+\frac{k_2 p^2}{4
   \langle k\rangle}-\frac{k_1 p}{4 \langle k\rangle}-\frac{k_2 p}{4
   \langle k\rangle}\\
&&-\frac{\cos \left(\frac{1}{2} \arg
   \left(c_1-ic_2 \right)\right)\sqrt[4]{c_2^2+c_1^2}}{12
   \langle k\rangle^2 \Gamma }\nonumber\\
\Re{\lambda_2}&=&\frac{k_1 p^2}{4
   \langle k\rangle}+\frac{k_2 p^2}{4 \langle k\rangle}-\frac{k_1
   p}{4 \langle k\rangle}-\frac{k_2 p}{4
   \langle k\rangle}\\
&&+\frac{\cos \left(\frac{1}{2} \arg
   \left(c_1-i c_2\right)\right)\sqrt[4]{c_2^2+c_1^2}}{12
   \langle k\rangle^2 \Gamma }\nonumber\\
\Re{\lambda_3}&=&\frac{k_1 p^2}{4
   \langle k\rangle}+\frac{k_2 p^2}{4 \langle k\rangle}-\frac{k_1
   p}{4 \langle k\rangle}-\frac{k_2 p}{4
   \langle k\rangle}-\\
&&\frac{ \cos \left(\frac{1}{2} \arg
   \left(c_1+i c_2\right)\right)\sqrt[4]{c_2^2+c_1^2}}{12
   \langle k\rangle^2 \Gamma }\nonumber\\
\Re{\lambda_4}&=&\frac{k_1 p^2}{4
   \langle k\rangle}+\frac{k_2 p^2}{4 \langle k\rangle}-\frac{k_1
   p}{4 \langle k\rangle}-\frac{k_2 p}{4
   \langle k\rangle}\\
&&+\frac{\cos \left(\frac{1}{2} \arg
   \left(c_1+i c_2 \right)\right)\sqrt[4]{c_2^2+c_1^2}}{12
   \langle k\rangle^2 \Gamma }.\nonumber
\end{eqnarray}
Here, $\arg(z)$ is the argument of the complex number $z$. The sum of the first four terms negative in each case because $0<p<1$ and for this reason $p^2<p$. The problem is the last summand, but it is possible to make a rough approximation of where all real parts are negative. As $\frac{\sqrt[4]{c_1^2+c_2^2}}{12\langle k\rangle^2\Gamma}>0$ and $-1\leq \cos(x)\geq 1$, all real parts are smaller than
\begin{eqnarray}
\frac{k_1 p^2}{4 \langle k\rangle}+\frac{k_2 p^2}{4
   \langle k\rangle}-\frac{k_1 p}{4 \langle k\rangle}-\frac{k_2 p}{4
   \langle k\rangle}+\frac{\sqrt[4]{c_2^2+c_1^2}}{12
   \langle k\rangle^2 \Gamma }
\end{eqnarray}
Inserting $c_1$ and $c_2$ and assuming $p\geq 0.5$ without loss of generality, one can approximate the fourth root in the last term by
\small
\begin{eqnarray}
\sqrt[4]{c_2^2+c_1^2}&=&\left(36\langle k\rangle^2(1-p)^2 p^2\Gamma^2\left(p^2k_1^3+(-2+p)\right.\right.\nonumber\\
&&\left.\left.pk_1^2k_2+(-1+p^2)k_1k_2^2+(1-p)^2k_2^3\right)^2\right.\nonumber\\
&&+\left.\left(-\langle k\rangle^2(1-p)^2p^2\Gamma^2(k_1+k_2)^2\right.\right.\nonumber\\
&&+\left.\left.\left(p^2k_1^2+(1-p)^2k_2^2\right)^2\right)^2\right)^\frac{1}{4}\nonumber\\
&<&\left(36\langle k\rangle^2(1-p)^2 p^2\Gamma^2\left(p^2k_1^3+(2+p)\right.\right.\nonumber\\
&&\left.\left.pk_1^2k_2+(1+p^2)k_1k_2^2+(1-p)^2k_2^3\right)^2\right.\nonumber\\
&&+\left.\left(\langle k\rangle^2(1-p)^2p^2\Gamma^2(k_1+k_2)^2\right.\right.\nonumber\\
&&+\left.\left.\left(p^2k_1^2+(1-p)^2k_2^2\right)^2\right)^2\right)^\frac{1}{4}\nonumber\\
&<&\left(36k_{max}^2p^4\Gamma^2(k_{max}^3(4p^2+2))^2\right.\nonumber\\
&&\left.+9(k_{max}^2p^4\Gamma^2(2k_{max})^2+(2p^2k_{max}^2)^2)^2\right)^\frac{1}{4}\nonumber\\
&<&\left(36\cdot 144\Gamma^2p^8k_{max}^8+(36\Gamma^2+4)^2p^8k_{max}^8\right)^\frac{1}{4}\nonumber\\
&=&p^2k_{max}^2\sqrt[4]{36\cdot144\Gamma^2+(36\Gamma^2+4)^2}
\end{eqnarray}
\normalsize
As the have checked numerically, the remaining root divided by $\sqrt[4]{36\cdot144\Gamma^2+(36\Gamma^2+4)^2}/\Gamma<9$ for $\omega<0.45$ (which is necessary to have mathematically meaningful probabilities between $0$ and $1$), so 
\begin{equation}
\frac{\sqrt[4]{c_2^2+c_1^2}}{12\langle k\rangle^2 \Gamma}<\frac{3p^2k_{max}^2}{4\langle k\rangle},
\end{equation}
and therefore all real parts are negative if
\begin{equation}
k_{max}+k_{min}>p\left(k_{max}+k_{min}+3\frac{k_{max}^2}{\langle k\rangle}\right), 
\end{equation}
or noted in a simplified way and remembering the assumption at the beginning of this approximation,
\begin{equation}
0.5\leq p<\frac{k_{max}+k_{min}}{k_{max}+k_{min}+3\frac{k_{max}^2}{\langle k\rangle}}.
\end{equation}.

\section{Theoretical approach: local update process on networks\label{local_theo}}
Similar as in the Moran process, in the local update process on networks the rate with which vertices with degree $k$ and strategy $\alpha$ carry this strategy over to nodes with degree $k^\prime$ and strategy $\alpha^\prime$ is given by
\begin{equation}
T_{k,k^\prime}^{\alpha^\prime\rightarrow\alpha}=\frac{p_{k^\prime}p_k k^\prime\rho_{\alpha,k}\rho_{\alpha^\prime,k^\prime}} {\langle k\rangle}\left(\frac{1}{2}+\frac{1}{2}\omega\frac{\pi_\alpha-\pi_{\alpha^\prime}} {\Delta\pi_{max}}\right),
\end{equation}
which does not vary from the results in well mixed populations in the case of homogenous degree distribution. The replicator equation for the model system of a heterogenous degree distribution with only two vertex degrees reads
\begin{eqnarray}
\label{eqB2}
\dot\rho_{0,k_1}&=&\frac{\psi k_1 p^2}{\langle k\rangle}\rho_{0,k_1}\left(\pi_0-\langle\pi\rangle_1\right)\\
&&+\frac{k_1 p(1-p)}{2\langle k\rangle}\left(\rho_{0,k_2}\left(1+\psi\left(\pi_0\langle\pi\rangle_1\right)\right)\right.\nonumber\\
&&\left.-\rho_{0,k_1}\left(1+\psi\left(\langle\pi\rangle_2-\pi_0\right)\right)\right)\label{hofbauersigmund_local_2Knotengrade}\nonumber
\end{eqnarray}
with $\psi=\omega/\Delta\pi_{max}$. 
The other equations are obtained analogously.
Here again the position of the fixed point is not changed compared to the well mixed case. But again it is no longer possible to absorb the selection strength $\omega$ by a rescaling of time.

A linear stability analysis leads to the matrix $\underline{\underline{A}}$
\small
\begin{equation}
\left(
\begin{array}{cccc}
 \frac{k_1 (p-1) p}{2 \langle k\rangle} &
   \frac{k_1^2 p^2 \psi }{3 \langle k\rangle^2} &
   -\frac{k_1 (p-1) p}{2 \langle k\rangle} &
   -\frac{k_1 k_2 (p-1) p \psi }{3
   \langle k\rangle^2} \\
 -\frac{k_1^2 p^2 \psi }{3 \langle k\rangle^2} &
   \frac{k_1 (p-1) p}{2 \langle k\rangle} &
   \frac{k_1 k_2 (p-1) p \psi }{3
   \langle k\rangle^2} & -\frac{k_1 (p-1) p}{2
   \langle k\rangle} \\
 -\frac{k_2 (p-1) p}{2 \langle k\rangle} &
   -\frac{k_1 k_2 (p-1) p \psi }{3
   \langle k\rangle^2} & \frac{k_2 (p-1) p}{2
   \langle k\rangle} & \frac{k_2^2 (p-1)^2 \psi }{3
   \langle k\rangle^2} \\
 \frac{k_1 k_2 (p-1) p \psi }{3
   \langle k\rangle^2} & -\frac{k_2 (p-1) p}{2
   \langle k\rangle} & -\frac{k_2^2 (p-1)^2 \psi }{3
   \langle k\rangle^2} & \frac{k_2 (p-1) p}{2
   \langle k\rangle}
\end{array}
\right).
\label{Stablitat_local_Matrix}
\end{equation}
\normalsize
The eigenvalues of this matrix are again given by
\begin{flalign}
\lambda_1=&-\frac{1}{12 \langle k\rangle^2}\left(-3 (k_1+k_2) \langle k\rangle
   (p-1) p\right.&\\
&\left.+2 i \left(k_2^2 (p-1)^2+k_1^2
   p^2\right) \psi +\sqrt{c_1-ic_2}\right)&\nonumber\\
\lambda_2=&\frac{1}{12 \langle k\rangle^2}\left(3
   (k_1+k_2) \langle k\rangle (p-1) p\right.&\\
&\left.-2 i
   \left(k_2^2 (p-1)^2+k_1^2 p^2\right)
   \psi +\sqrt{c_1-ic_2}\right)&\nonumber
\end{flalign}
\begin{flalign}
\lambda_3=&\frac{1}{12 \langle k\rangle^2}\left(2 i k_1^2
   \psi  p^2+3 k_1 \langle k\rangle (p-1) p\right.&\\
&\left.+k_2
   (p-1) (3 \langle k\rangle p+2 i k_2 (p-1) \psi
   )-\sqrt{c_1+ic_2}\right)&\nonumber\\
\lambda_4=&\frac{1}{12 \langle k\rangle^2}\left(2 i k_1^2
   \psi  p^2+3 k_1 \langle k\rangle (p-1) p\right.&\\
&\left.+k_2
   (p-1) (3 \langle k\rangle p+2 i k_2 (p-1) \psi
   )+\sqrt{c_1+ic_2}\right)&\nonumber
\end{flalign}
with the abbreviations
\begin{eqnarray}
c_1&=&9 (k_1+k_2)^2
   \langle k\rangle^2 (p-1)^2 p^2\\
&&-4
   \left(k_2^2 (p-1)^2+k_1^2 p^2\right)^2
   \psi ^2\nonumber\\
c_2&=&12 \langle k\rangle (p-1) \left(p^2
   k_1^3+k_2 (p-2) p
   k_1^2\right.\\
&&\left.+k_2^2 \left(p^2-1\right)
   k_1+k_2^3 (p-1)^2\right) \psi  p.
\nonumber
\end{eqnarray}
The real parts of this eigenvalues are given by
\begin{eqnarray}
\Re\lambda_1&=&\frac{k_1 p^2}{4
   \langle k\rangle}+\frac{k_2 p^2}{4
   \langle k\rangle}-\frac{k_1 p}{4
   \langle k\rangle}-\frac{k_2 p}{4
   \langle k\rangle}\\
&&-\frac{\cos \left(\frac{1}{2} \arg \left(c_1-ic_2\right)\right)\sqrt[4]{c_1^2+c_2^2}}{12
   \langle k\rangle^2}\nonumber\\
\Re\lambda_2&=&\frac{k_1 p^2}{4
   \langle k\rangle}+\frac{k_2 p^2}{4
   \langle k\rangle}-\frac{k_1 p}{4
   \langle k\rangle}-\frac{k_2 p}{4
   \langle k\rangle}\\
&&+\frac{\cos \left(\frac{1}{2} \arg \left(c_1-ic_2\right)\right)\sqrt[4]{c_1^2+c_2^2}}{12
   \langle k\rangle^2}\nonumber\\
\Re\lambda_3&=&\frac{k_1 p^2}{4
   \langle k\rangle}+\frac{k_2 p^2}{4
   \langle k\rangle}-\frac{k_1 p}{4
   \langle k\rangle}-\frac{k_2 p}{4
   \langle k\rangle}\\
&&-\frac{\cos \left(\frac{1}{2} \arg \left(c_1+ic_2\right)\right)\sqrt[4]{c_1^2+c_2^2}}{12
   \langle k\rangle^2}\nonumber\\
\Re\lambda_4&=&\frac{k_1 p^2}{4
   \langle k\rangle}+\frac{k_2 p^2}{4
   \langle k\rangle}-\frac{k_1 p}{4
   \langle k\rangle}-\frac{k_2 p}{4
   \langle k\rangle}\\
&&+\frac{\cos \left(\frac{1}{2} \arg \left(c_1+ic_2\right)\right)\sqrt[4]{c_1^2+c_2^2}}{12 \langle k\rangle^2}\nonumber,
\end{eqnarray}
which is very similar to the corresponding result in the Moran process (it varies only slightly in $c_1$ and $c_2$). Searching for an upper bound for the real parts of the eigenvalues, we even derive the same inequality as for the Moran process
\begin{equation}
 0.5\leq p<\frac{k_{max}+k_{min}}{k_{max}+k_{min}+3\frac{k_{max}^2}{\langle k\rangle}}.
\end{equation}
%\mbox{}\\ \vspace{6in}

%\clearpage

\end{document}